\documentclass[journal,]{IEEEtran}

\usepackage{graphicx}
\usepackage{epstopdf}
\usepackage{standalone}
\usepackage[nolist ]{acronym}
\usepackage{tcolorbox}

\usepackage{xstring}

\usepackage[utf8]{inputenc}
\usepackage{marvosym}

\usepackage{algorithm}
\usepackage{algpseudocode}
\usepackage{tikz}
\usetikzlibrary{positioning}

\usepackage{pgfplots}
\pgfplotsset{compat=newest}
\pgfplotsset{
    box plot/.style={
        /pgfplots/.cd,
        fill=blue!30,
        only marks,
        mark=-,
        mark size=0.2em,
        /pgfplots/error bars/.cd,
        y dir=plus,
        y explicit,
    },
    box plot box/.style={
        /pgfplots/error bars/draw error bar/.code 2 args={%
            \draw  ##1 -- ++(.2em,0pt) |- ##2 -- ++(-.2em,0pt) |- ##1 -- cycle;
        },
        /pgfplots/table/.cd,
        y index=2,
        y error expr={\thisrowno{3}-\thisrowno{2}},
        /pgfplots/box plot
    },
    box plot top whisker/.style={
        /pgfplots/error bars/draw error bar/.code 2 args={%
            \pgfkeysgetvalue{/pgfplots/error bars/error mark}%
            {\pgfplotserrorbarsmark}%
            \pgfkeysgetvalue{/pgfplots/error bars/error mark options}%
            {\pgfplotserrorbarsmarkopts}%
            \path ##1 -- ##2;
        },
        /pgfplots/table/.cd,
        y index=4,
        y error expr={\thisrowno{2}-\thisrowno{4}},
        /pgfplots/box plot
    },
    box plot bottom whisker/.style={
        /pgfplots/error bars/draw error bar/.code 2 args={%
            \pgfkeysgetvalue{/pgfplots/error bars/error mark}%
            {\pgfplotserrorbarsmark}%
            \pgfkeysgetvalue{/pgfplots/error bars/error mark options}%
            {\pgfplotserrorbarsmarkopts}%
            \path ##1 -- ##2;
        },
        /pgfplots/table/.cd,
        y index=5,
        y error expr={\thisrowno{3}-\thisrowno{5}},
        /pgfplots/box plot
    },
    box plot median/.style={
        /pgfplots/box plot
    },
    boxplot/every median/.style={
    	ultra thick,dashed,cyan
    }
}

\definecolor{flexicolor}{RGB}{46,49,146}
\definecolor{amaricolor}{RGB}{237,28,36}

\usepackage{xspace}

\usepackage[binary-units=true]{siunitx}
\usepackage{psfrag}
\usepackage{graphicx}
\usepackage{tabularx,booktabs}
\usepackage{multirow}
\usepackage{rotating}

\usepackage{wrapfig}

\usepackage{multirow}

\usepackage[cmex10]{amsmath}
\usepackage{upgreek}

\usepackage[caption=false,font=footnotesize]{subfig}
\usepackage{stfloats}
\begin{document}

\newcommand{\paperTitle}{Boosting Vehicle-to-cloud Communication by Machine Learning-enabled Context Prediction}
\newcommand{\githubUrl}{\footnote{Available at https://github.com/BenSliwa/MTCApp}\xspace}

\newcommand{\figurePadding}{0pt}
\newcommand{\figureTopPadding}{\figurePadding}
\newcommand{\figureBottomPadding}{\figurePadding}

\ifdefined\singleColumn
	\newcommand{\single}{1\textwidth}
	\newcommand{\dual}{.48\textwidth}
	\newcommand{\triple}{.31\textwidth}
	\newcommand{\singleC}{0.5\columnwidth}
	\newcommand{\BLS}{2}
\else
	\newcommand{\single}{1\textwidth}
	\newcommand{\dual}{.48\textwidth}
	\newcommand{\triple}{.32\textwidth}
	\newcommand{\singleC}{1\columnwidth}
	\newcommand{\BLS}{1}
\fi

\newcommand\vs[1]
{
	\ifdefined\singleColumn
		\vspace{#1}
	\else

	\fi
}

\newcommand{\entry}{}
\newcommand{\head}{}

\newcommand{\dummy}[3]
{
	\begin{figure}[b!]  
		\begin{tikzpicture}
		\node[draw,minimum height=6cm,minimum width=\columnwidth]{\LARGE #1};
		\end{tikzpicture}
		\caption{#2}
		\label{#3}
	\end{figure}
}

\newcommand{\basicFig}[7]
{
	\begin{figure}[#1]  	
		\vspace{#6}
		\centering

		\ifdefined\singleColumn
		\includegraphics[width=0.6\columnwidth]{#2}
		\else
		\includegraphics[width=#7\columnwidth]{#2}
		\fi
		
		\caption{#3}
		\label{#4}
		\vspace{#5}	
	\end{figure}
}
\newcommand{\fig}[4]{\basicFig{#1}{#2}{#3}{#4}{0cm}{0cm}{1}}

\newcommand{\subfig}[3]
{
	\subfloat[#3]
	{
		\includegraphics[width=#2\textwidth]{#1}
	}
	\hfill
}

\newcommand\circled[1] 
{
	\tikz[baseline=(char.base)]
	{
		\node[shape=circle,draw,inner sep=1pt, outer sep=0](char){#1};
	}\xspace
}


\renewcommand{\vec}{\mathbf}					
\newcommand{\vecg}{\boldsymbol}		 
\renewcommand{\tilde}{\widetilde}

\newcommand{\bx}{\boldsymbol{x}}
\newcommand{\bt}{\boldsymbol{\beta}}
\newcommand{\cD}{\mathcal{D}}
\newcommand{\cT}{\mathcal{T}}
\newcommand{\cO}{\mathcal{O}}
\newcommand{\cE}{\mathcal{E}}
\newcommand{\cM}{\mathcal{M}}
\newcommand{\mR}{\mathbb{R}}
\newcommand{\mP}{\mathbb{P}}
\begin{acronym}
	\acro{CAT}{Channel-aware Transmission}
	\acro{CoPoMo}{Context-aware Power Consumption Model}
	\acro{CPS}{Cyber Physical System}
	\acro{IoT}{Internet of Things}	
	\acro{pCAT}{predictive CAT}
	\acro{ML-CAT}{Machine Learning CAT}
	\acro{ML-pCAT}{Machine Learning pCAT}
	\acro{MTC}{Machine-type Communication}
	\acro{SDR}{Software-defined Radio}
	\acro{MANET}{Mobile Ad-hoc Network}
	\acro{V2V}{Vehicle-to-vehicle}
	\acro{V2X}{Vehicle-to-everything}
	\acro{UAV}{Unmanned Aerial Vehicle}
	\acro{RSU}{Roadside Unit}
	\acro{CR}{Cognitive Radio}
	\acro{REM}{Radio Environment Map}
	
	\acro{LTE}{Long Term Evolution}
	\acro{mmWave}{millimeter wave}
	
	\acro{M2M}{Machine-to-machine}
	\acro{H2H}{Human-to-human}
	\acro{ITS}{Intelligent Transportation System}
	\acro{SCATS}{Sydney Coordinated Adaptive Traffic System}
	\acro{KPI}{Key Performance Indicator}

	%
	%
	\acro{UE}{User Equipment}
	\acro{PRB}{Physical Resource Block}
	\acro{FPC}{Fractional Path Loss Compensation}
	\acro{MCS}{Modulation and Coding Scheme}

	%
	%
	\acro{ARIMA}{AutoRegressive Integrated and Moving Average}
	\acro{MAE}{Mean Absolute Error}
	\acro{RMSE}{Root Mean Square Error}
	\acro{AoI}{Age of Information}
	\acro{RSSI}{Received Signal Strength Indicator}
	\acro{RSSNR}{Reference Signal SNR}
	\acro{RB}{Resource Block}
	\acro{KPI}{Key Performance Indicator}
	\acro{MAC}{Medium Access Control}
	\acro{PFR}{Prediction Failure Ratio}
	
	%
	%
	\acro{ANN}{Artificial Neural Network}
	\acro{M5T}{M5 Decision Tree}
	\acro{SVM}{Support Vector Machine}
	\acro{LR}{Linear Regression}
	\acro{RR}{Ridge Regression}
	\acro{RF}{Random Forest}
	\acro{DL}{Deep Learning}
	
	%
	%
	\acro{CQI}{Channel Quality Indicator}
	\acro{RSRP}{Reference Signal Received Power}
	\acro{RSRQ}{Reference Signal Received Quality}
	\acro{SNR}{Signal-to-noise Ratio}
	\acro{SINR}{Signal-to-interference-plus-noise Ratio}
	\acro{RS-SINR}{Reference Signal Signal to Interference and Noise Ratio}
	
	\acro{HTTP}{Hypertext Transfer Protocol}
	\acro{TCP}{Transmission Control Protocol}
	
	\acro{GPS}{Global Positioning System}
	\acro{WGS84}{World Geodetic System 1984}
	
\end{acronym}

\newcommand{\dBm}{dBm}
\newcommand{\dBi}{dBi}

\newcommand{\cf}{cf.\xspace}		
\newcommand{\eg}{e.g.\xspace}		
\newcommand{\ie}{i.e.\xspace}		

\newcommand{\Idle}{\textsf{\small{}Idle}\xspace}
\newcommand{\Low}{\textsf{\small{}Low}\xspace}
\newcommand{\High}{\textsf{\small{}High}\xspace}
\newcommand{\Max}{\textsf{\small{}Max}\xspace}

\newcommand{\Fig}[1]{Fig.~#1\xspace}
\renewcommand{\Sec}[1]{Sec.~#1\xspace}
\newcommand{\Tab}[1]{Tab.~#1\xspace}
\newcommand{\Eq}[1]{Eq.~#1\xspace}

\newcommand{\hex}[1]{\texttt{0x#1}\xspace}

\newcommand{\UE}{\ac{UE}\xspace}
\newcommand{\UEs}{\acp{UE}\xspace}
\newcommand{\LTE}{\ac{LTE}\xspace}
\newcommand{\SNR}{\ac{SNR}\xspace}
\newcommand{\SINR}{\ac{SINR}\xspace}
\newcommand{\RSSINR}{\ac{RS-SINR}\xspace}
\newcommand{\RSRP}{\ac{RSRP}\xspace}
\newcommand{\RSRQ}{\ac{RSRQ}\xspace}
\newcommand{\TCP}{\ac{TCP}\xspace}

\acrodef{FPC}{Fractional Path Loss Compensation}
\newcommand{\PFC}{\ac{PFC}\xspace}
\acrodef{eNB}{evolved NodeB}
\newcommand{\eNB}{\ac{eNB}\xspace}
\newcommand{\GPS}{\ac{GPS}\xspace}
\acrodef{CPU}{Central Processing Unit}
\newcommand{\CPU}{\ac{CPU}\xspace}
\acrodef{SoC}{System on a Chip}
\newcommand{\SoC}{\ac{SoC}\xspace}
\acrodef{COTS}{Commercial Off-the-Shelf}
\newcommand{\COTS}{\ac{COTS}\xspace}
\acrodef{MCS}[MCS]{Modulation and Coding Scheme}
\newcommand{\MCS}{\ac{MCS}\xspace}
\acrodef{TBS}[TBS]{Transport Block Size}
\newcommand{\TBS}{\ac{TBS}\xspace}
\newcommand{\PRB}{\ac{PRB}\xspace}
\newcommand{\PRBs}{\acp{PRB}\xspace}
\acrodef{FPC}{Fractional Path Loss Compensation}
\newcommand{\FPC}{\ac{FPC}\xspace}
\acrodef{TPC}{Transmission Power Control}
\newcommand{\TPC}{\ac{TPC}\xspace}
\acrodef{PDCCH}{Physical Downlink Control Channel}
\newcommand{\PDCCH}{\ac{PDCCH}\xspace}
\acresetall
\title{\paperTitle}

\author{Benjamin Sliwa$^1$, Robert Falkenberg$^1$, Thomas Liebig$^2$, Nico Piatkowski$^2$, and Christian Wietfeld$^1$%
	\thanks{$^1$Benjamin Sliwa, Robert Falkenberg and Christian Wietfeld are with Communication Networks Institute, TU Dortmund University, 44227 Dortmund, Germany
		{\tt\small $\{$Benjamin.Sliwa, Robert.Falkenberg, Christian.Wietfeld$\}$@tu-dortmund.de}}%
	\thanks{$^2$Thomas Liebig and Nico Piatkowski are with the Computer Science Department, AI Group, TU Dortmund University, 44227 Dortmund, Germany
		{\tt\small $\{$Nico.Piatkowski, Thomas.Liebig$\}$@tu-dortmund.de}}%
}

\maketitle
%
%
\def\COPYRIGHTYEAR{2019}
\def\CONFERENCE{IEEE Transactions on Intelligent Transportation Systems}
\ifx\CONFERENCE\VOID
\def\conferencenotice{Submitted for publication}
\def\copyrightnotice{}
\else
\ifx\DOI\VOID
\def\conferencenotice{Accepted for publication in: \CONFERENCE}	
\else
\def\conferencenotice{Published in: \CONFERENCE\\DOI: \href{http://dx.doi.org/\DOI}{\DOI}
	
	\vspace{0.3cm}
	\pdfcomment[color=yellow,icon=Note]{\bibtex}    
	
}
\fi
\def\copyrightnotice{
	\copyright~\COPYRIGHTYEAR~IEEE. Personal use of this material is permitted. Permission from IEEE must be obtained for all other uses, including reprinting/republishing this material for advertising or promotional purposes, collecting new collected works for resale or redistribution to servers or lists, or reuse of any copyrighted component of this work in other works.
}
\fi
\def\overlayimage{%
	\begin{tikzpicture}[remember picture, overlay]
	\node[below=5mm of current page.north, text width=20cm,font=\sffamily\footnotesize,align=center] {\conferencenotice};
	\node[above=5mm of current page.south, text width=15cm,font=\sffamily\footnotesize] {\copyrightnotice};
	\end{tikzpicture}%
}
\overlayimage
\begin{abstract}
		
The exploitation of vehicles as mobile sensors acts as a catalyst for novel crowdsensing-based applications such as intelligent traffic control and distributed weather forecast. However, the massive increases in \ac{MTC} highly stress the capacities of the network infrastructure.
%
%
With the system-immanent limitation of resources in cellular networks and the resource competition between human cell users and \ac{MTC}, more resource-efficient channel access methods are required in order to improve the coexistence of the different communicating entities.
%
%
In this paper, we present a machine learning-enabled transmission scheme for client-side opportunistic data transmission. By considering the measured channel state as well as the predicted future channel behavior, delay-tolerant \ac{MTC} is performed with respect to the anticipated resource-efficiency.
%
%
The proposed mechanism is evaluated in comprehensive field evaluations in public \ac{LTE} networks, where it is able to increase the mean data rate by 194\% while simultaneously reducing the average power consumption by up to 54\%.

\end{abstract}

\begin{IEEEkeywords}
	Context-predictive Communication, Machine Learning, Crowdsensing, Intelligent Transportation Systems, Mobile Sensors
\end{IEEEkeywords}

\IEEEpeerreviewmaketitle

\section{Introduction} \label{sec:introduction}

%
%
While cars were only seen as means for personal transportation in the past, they are currently transcending to mobile sensor nodes that provide crowdsensing-based services with highly up-to-date information \cite{Cavalcanti/etal/2018a}. Applications range from predictive maintenance and intelligent traffic control to road-roughness detection \cite{Wang/etal/2017a} and distributed weather forecast \cite{Calafate/etal/2017a}. In addition, small-scale autonomous robots such as \acp{UAV} are expected to become native parts of \acp{ITS} \cite{Menouar/etal/2017a}. Since the operation time of these vehicles is highly determined by the available energy resources, energy-efficient communication has become one of the major research fields in mobile robotics \cite{Zeng/Zhang/2017a}.
%
%
Different communication technologies with characteristic system properties and intended use-cases are currently being investigated for interconnecting vehicles and infrastructure. Among others, the framework for the deployment of \acp{ITS} of the European Parliament \cite{EU/2019a} proposes the usage IEEE 802.11p-based ad-hoc communication for safety-related \ac{V2V} data transfer with low latencies and small data packets. However, this technology is not able to provide internet-based vehicle-to-cloud connectivity, as there are practically no deployments of \acp{RSU}, which offer the required gateway functionalities. Therefore, delay-tolerant and data-intense messaging is intended to be carried out based on existing cellular communication technologies (e.g., \ac{LTE} and upcoming 5G networks), which already offer large-scale coverage.
With the expected massive increase in vehicular \ac{MTC} \cite{Djahel/etal/2015a} and the general growth of cellular data traffic \cite{Zhang/Arvidsson/2012a}, the network infrastructure is facing the challenge of resource-competition between human cell users and \ac{IoT}-related data transmissions \cite{Zanella/etal/2014a}. Fig.~\ref{fig:application_scenario} gives an overview about the requirements of different vehicular and \ac{IoT}-based communication systems and the resulting challenges that arise from the channel dynamics and the limited cell resources.
%
%
\basicFig{b}{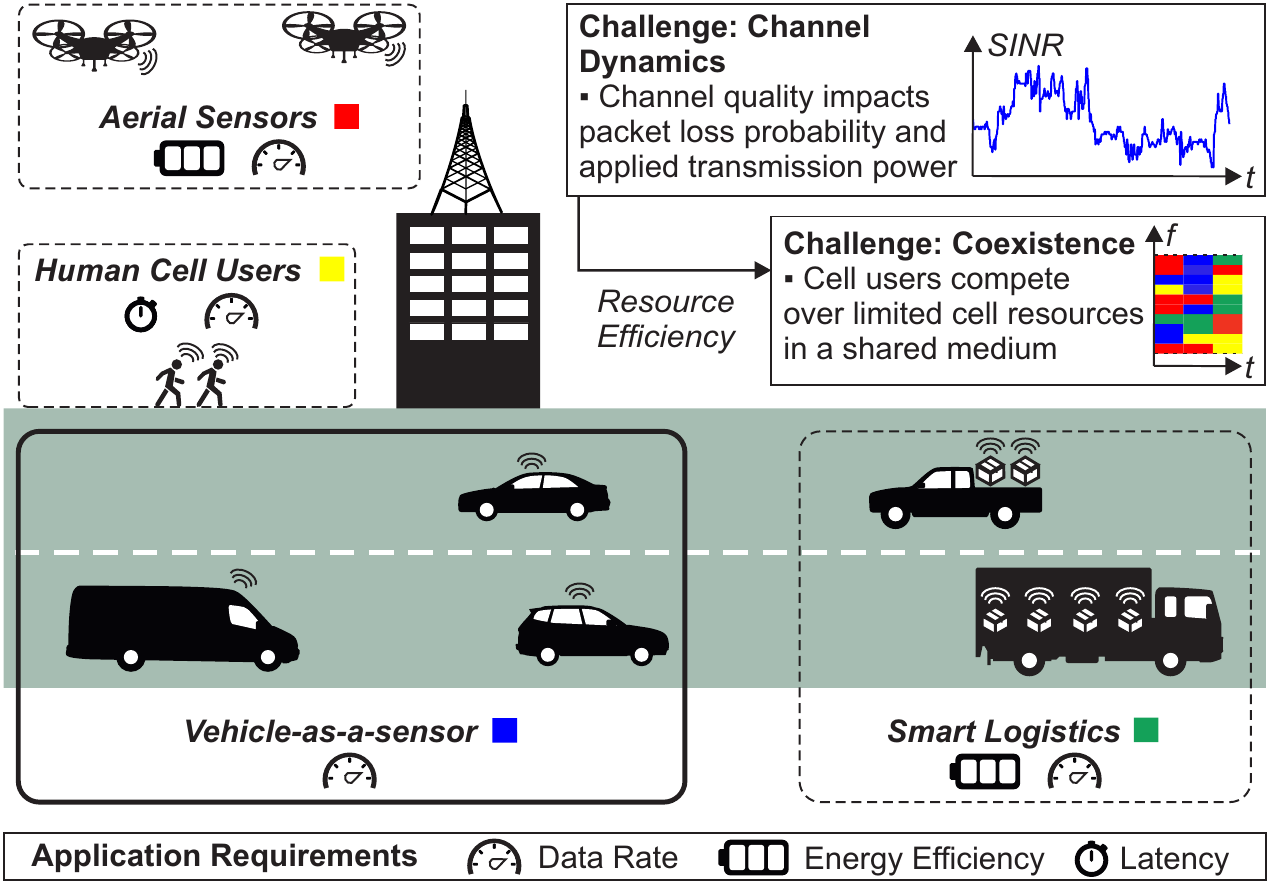}{Challenges and application-specific requirements of \ac{ITS}-based data transmission. In the paper, we exploit knowledge about the channel dynamics for opportunistic data transfer in oder to avoid wastage of cell resource. Although the empirical evaluation focuses on the vehicle-as-a-sensor use case, the results are also relevant for smart logistics and aerial sensing.}{fig:application_scenario}{0cm}{-0.5cm}{1}
%
%
%

%
%
A promising approach to address these issues is the application of \emph{context-aware communication} \cite{Bui/etal/2017a} that exploits the dynamics of the communication channel to schedule delay-tolerant transmissions in an \emph{opportunistic} way for increasing the transmission efficiency with regard to data rate, packet loss probability and energy consumption. As a consequence, communication resources are occupied for shorter time intervals and can early be used by other cell users, which enables a better coexistence and overall system performance \cite{Ide/etal/2015a}.

%
%
In this paper, we extend and bring together the methods, results and insights of previous work \cite{Wietfeld/etal/2014a,Pillmann/etal/2017a,Sliwa/etal/2018a,Sliwa/etal/2018b,Falkenberg/etal/2018a} on context-aware car-to-cloud communication and propose a client-side opportunistic transmission scheme that applies machine learning-based data rate prediction for scheduling the transmission times of sensor data transmissions with respect to the expected resource-efficiency. Moreover, mobility prediction and connectivity maps are exploited to integrate the anticipated future channel behavior into the transmission process. The analysis focuses on resource-aware machine learning models that allow the online-prediction on off-the-shelf smartphones and embedded systems without causing significant additional computation overheads themselves.

The analysis in this work focuses on exploiting vehicles as \emph{moving sensor nodes} that transmit measured data to cloud-based services through a cellular network. It is assumed that the actual data-driven crowdsensing applications specify \emph{soft deadlines}, which allow the opportunistic and delay-tolerant approach of the proposed transmission scheme.
Examples for existing real-world systems with similar requirements are \ac{SCATS}-based ramp metering in \acp{ITS} (traffic flow is communicated every six minutes), real-time traffic optimization \cite{Shi/Abdel-Aty/2015a,Vandenberghe/etal/2012a} (sensors provide their information once per minute) and air pollution monitoring \cite{Chen/etal/2018a}, which is performed every five minutes.

The key contributions of this paper are the following:
\begin{itemize}
	\item A \textbf{highly configurable probabilistic model} for opportunistic vehicle-to-cloud data transfer with respect to the channel properties.
	\item Machine learning-based \textbf{uplink data rate prediction} based on measured passive downlink indicators, which is applied as a metric to schedule data transmissions.
	\item \textbf{Mobility prediction} using navigation system knowledge to forecast the future vehicle position and allow exploitation of a priori information about the transmission channel by using multi-layer connectivity maps.	
	\item A \textbf{closed process for post-processing uplink power consumption analysis} exploiting machine learning-based transmission power prediction from passive downlink indicators and device-specific laboratory measurements.
	\item The developed measurement applications and the raw measurement results are provided as \textbf{Open Source}.
\end{itemize}
%
%
The remainder of the manuscript is structured as follows. After discussing relevant state-of-the-art approaches in \Sec{\ref{sec:related_work}}, the proposed transmission scheme and its individual components are presented in \Sec{\ref{sec:approach}}. Afterwards, a machine learning-empowered process to assess the uplink power consumption of embedded devices is presented in \Sec{\ref{sec:tx_power}}. \Sec{\ref{sec:methodoloy}} gives an overview about the methodological setup for the field evaluation and finally, the achieved results of the different models are evaluated and discussed in \Sec{\ref{sec:results}}. 
\\

\vspace{-0.5cm}
\section{Related Work} \label{sec:related_work}

%
%
A detailed evaluation about the complex interdependencies of \ac{M2M} and \ac{H2H} data traffic as well as their coexistence in the same cellular network is performed in \cite{Ide/etal/2015a}. The optimization of the coexistence of these data traffic types is often addressed on the network infrastructure side, e.g., by cognitive and channel-dependent scheduling mechanisms \cite{Giluka/etal/2018a} that consider different traffic types and priorities. In \cite{Feng/etal/2017a}, the authors propose a biology-inspired approach that considers \ac{M2M} and \ac{H2H} as populations of predators and prey in order to achieve a stable equilibrium for both traffic types.
Although these optimizations might have an impact on the design of future networks, they can often not be applied within existing networks as the involved changes could lead to incompatibilities. Moreover, the scientific evaluation of infrastructure-side optimizations is often limited to simulation scenarios due to lacking access to the required hardware equipment and the inherent complexity of real-world scenarios.

%
%
Especially for platooning, one way to reduce the crowdsensing-related cell load is to pre-aggregate the sensor data in a gateway vehicle \cite{Hu/etal/2016a} before it is actually transmitted in order to avoid the transfer of redundant information. Alternative approaches are provided by \emph{social-based forwarding} \cite{Hui/etal/2011a} and offloading techniques \cite{Li/etal/2014a}. Within this paper, we focus on optimizing the transmission behavior of individual non-coordinated vehicles.

%
%
Anticipatory mobile networking aims to raise the situation-awareness of the communicating entities by integrating additional information into the decision processes in order to optimize different individual \acp{KPI} or the overall system performance \cite{Wan/etal/2014a}. The \emph{anticipatory communication paradigm} is closely related to the application of machine learning, which can be exploited for the prediction of future behaviors and the consideration of hidden parameters that are not directly accessible within the complex system \cite{Jiang/etal/2017a}. In \cite{Bui/Widmer/2018a}, the authors propose a data-driven framework for optimizing the resource-efficiency of the network infrastructure by centralized and distributed predictions using control channel analysis. Within the case-study, about 95\% of the overall traffic value was precisely predicted, which enabled the network operators to more than double the offered data rate using the optimization framework. 
While theoretically, the resulting data rate of a transmission is the result of a deterministic process, predicting those values proactively within a live-system is a challenging task due to the large number of involved hidden influences (e.g., scheduling, packet loss, channel stability, spectrum sharing and cross-layer interdependencies) \cite{Falkenberg/etal/2017b}. 
%
%
Different authors have investigated the impact of the channel quality to the resulting data rate of cellular data transmissions \cite{Sliwa/etal/2018b, Jomrich/etal/2018a, Samba/etal/2017a, Sliwa/Wietfeld/2019b, Akselrod/etal/2017a} in different environments that range from highway to inner city scenarios. The studies agree that \emph{passively} measurable network quality indicators such as  \RSRP, \RSRQ, \SINR, and \ac{CQI} provide meaningful information, which can be leveraged to estimate the resulting data rate based on machine learning methods even in challenging environments. In comparison to time series-based \emph{active} data rate prediction (e.g., based on Kalman filters), passive approaches do not monitor the data rates of ongoing transmissions and can therefore be applied without introducing additional traffic themselves. As resource efficiency is one of the optimization goals of this work, we focus on passive data rate prediction.

%
%
For mobile wireless networks, the dynamics of the communication channel are highly affected by the mobility characteristics of the moving vehicle \cite{Chen/etal/2017a}. Therefore, mobility-awareness allows the explicit consideration of these impact factors e.g., for handover optimization \cite{Cia/etal/2017a} and improved routing in vehicular ad-hoc networks \cite{Sliwa/etal/2016a}.

%
%
While crowdsensing forms the considered application scenario for this work and provides the reason for the actual vehicle-to-cloud data transfer, the technique itself can be exploited in order to optimize the environmental awareness of the vehicles. In this context, the usage of \emph{connectivity maps} for anticipatory communication \cite{Kasparick/etal/2016a,Poegel/Wolf/2015a} allows to exploit a priori information about the channel quality based on previous measurements in the same geographical area. \acp{REM} implement a similar concept, which enables opportunistic data transfer with \ac{CR} methods \cite{Haykin/2005a}. However, those purely spectrum-aware approaches do not consider the cross-layer interdependencies within the protocol stack. Moreover, as the resource allocation in \ac{LTE} is performed by the scheduling mechanisms of the \ac{eNB}, those methods have to be implemented by the mobile network operator. In contrast to that, the proposed machine learning-based approach can easily be implemented on the client side without requiring modifications to the network infrastructure.

\section{Machine Learning-enabled Transmission of Vehicular Sensor Data} \label{sec:approach}


In this section, the machine learning-based sensor data transmission schemes and their corresponding components are presented. In the first step, the legacy \ac{CAT} scheme \cite{Ide/etal/2015a} is generalized and augmented using machine learning-based data rate prediction. Afterwards, we transit from context-aware to context-predictive communication with the extended \ac{pCAT} that exploits multi-layer connectivity maps and mobility prediction to consider the anticipated future network state in the transmission process. Finally, the main contributions of this paper --- the transmission schemes \textbf{\ac{ML-CAT}} and \textbf{\ac{ML-pCAT}} --- are derived by bringing the key insights together.
%
%
\begin{figure*}[] 
	\centering
	\includegraphics[width=\single]{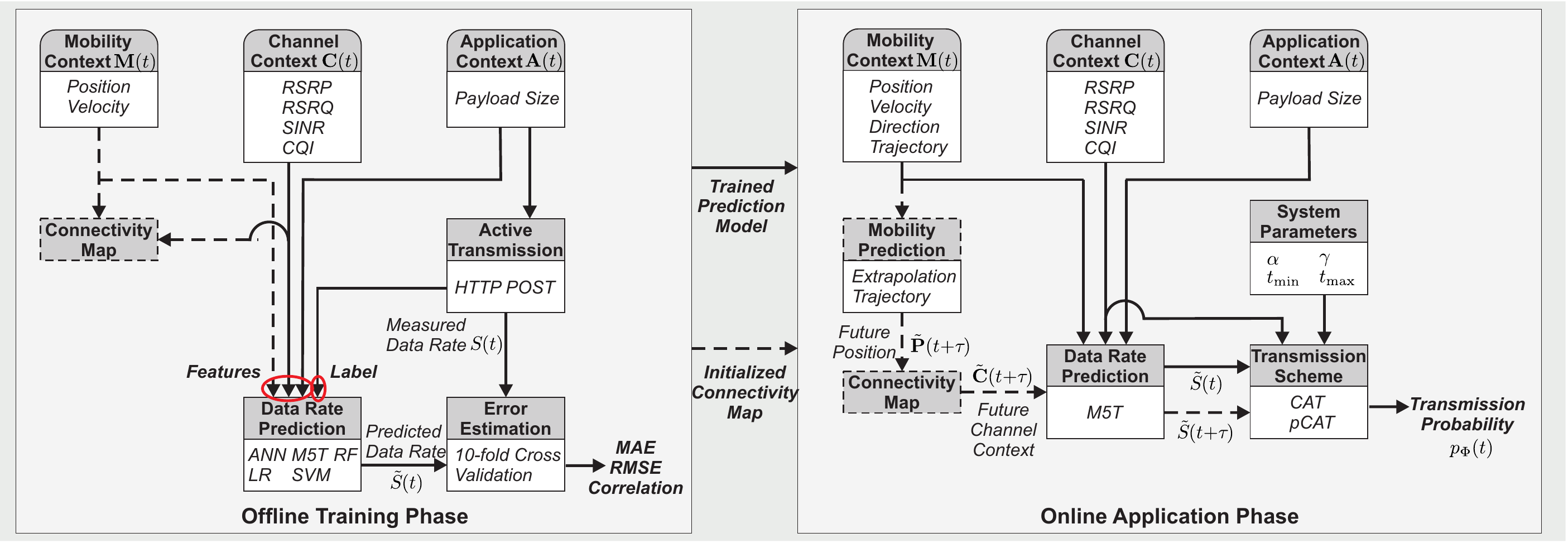}
	\vspace{-0.6cm}
	\caption{Overall system architecture model for training and application phase of the probabilistic transmission scheme. The dashed components are only required for the context-predictive \ac{pCAT}-based transmissions. All modules operate on the application layer.}
	\label{fig:architecture}
	\vspace{-0.5cm}
\end{figure*}
The overall system architecture model of the proposed approach that operates on the application layer is shown in \Fig{\ref{fig:architecture}}. Transmissions are performed \emph{probabilistically} with respect to the network quality by exploiting \emph{connectivity hotspots} that allow fast and reliable data delivery and avoiding \emph{connectivity valleys} that implicate high packet loss probabilities. The acquired sensor data is stored in a local buffer until a transmission decision has been made for the whole data buffer.
\\
%
%
The \textbf{training phase} consists of \emph{passive probing} of the \ac{LTE} downlink indicators --- which form the \emph{channel context} $\vec{C}(t)$ --- as well as of active data transmission with variable payload sizes using \ac{HTTP} POST.
%
%
The \emph{feature set} of the data rate prediction is composed of the network quality indicators, the velocity and the payload size of the data packet. The resulting data rate of the active transmission is used as the \emph{label} for the prediction process, which is performed with the models \emph{\ac{ANN}}, \emph{\ac{LR}}, \emph{\ac{RF}}, \emph{\ac{M5T}} and \emph{\ac{SVM}}. Finally, the prediction performance of the different models is evaluated using 10-fold cross validation.
%
%
Additionally, the measured \emph{channel context} parameters and the position information of the vehicle are utilized to create a \emph{multi-layer connectivity map} that stores the cell-wise average of each indicator from multiple visits of the same geographical area.
\\
%
%
During the \textbf{application phase}, the context information is leveraged to calculate the channel-aware transmission probability $p_{\Phi}(t)$. The channel is only probed passively and the most accurate previously trained prediction model uses measurements for the \emph{channel context} $\vec{C}(t)$, the \emph{mobility context} $\vec{M}(t)$ and the \emph{application context} $\vec{A}(t)$ to predict the currently achievable data rate $\tilde{S}(t)$.
The latter forms the metric for the transmission scheme that is configured with multiple \emph{system parameters} (see \Sec{\ref{sec:cat_tx}}). Alternatively, the channel quality indicators can be used directly to serve as a transmission metric.
\\
%
%
For the context-predictive \ac{pCAT} transmission scheme  (see \Sec{\ref{sec:pcat_tx}}), \emph{mobility prediction} is applied to estimate the future position $\tilde{\vec{P}}(t+\tau)$ for a defined prediction horizon $\tau$. $\tilde{\vec{P}}(t+\tau)$ is then used to access the corresponding cell entry in the connectivity map in order to obtain an estimation for the future channel context $\tilde{\vec{C}}(t+\tau)$. This knowledge about the \emph{anticipated channel behavior} enables the calculation of the forecasted data rate $\tilde{S}(t+\tau)$ that is integrated into the transmission process of the proposed \ac{pCAT}.
%
%
\basicFig{b}{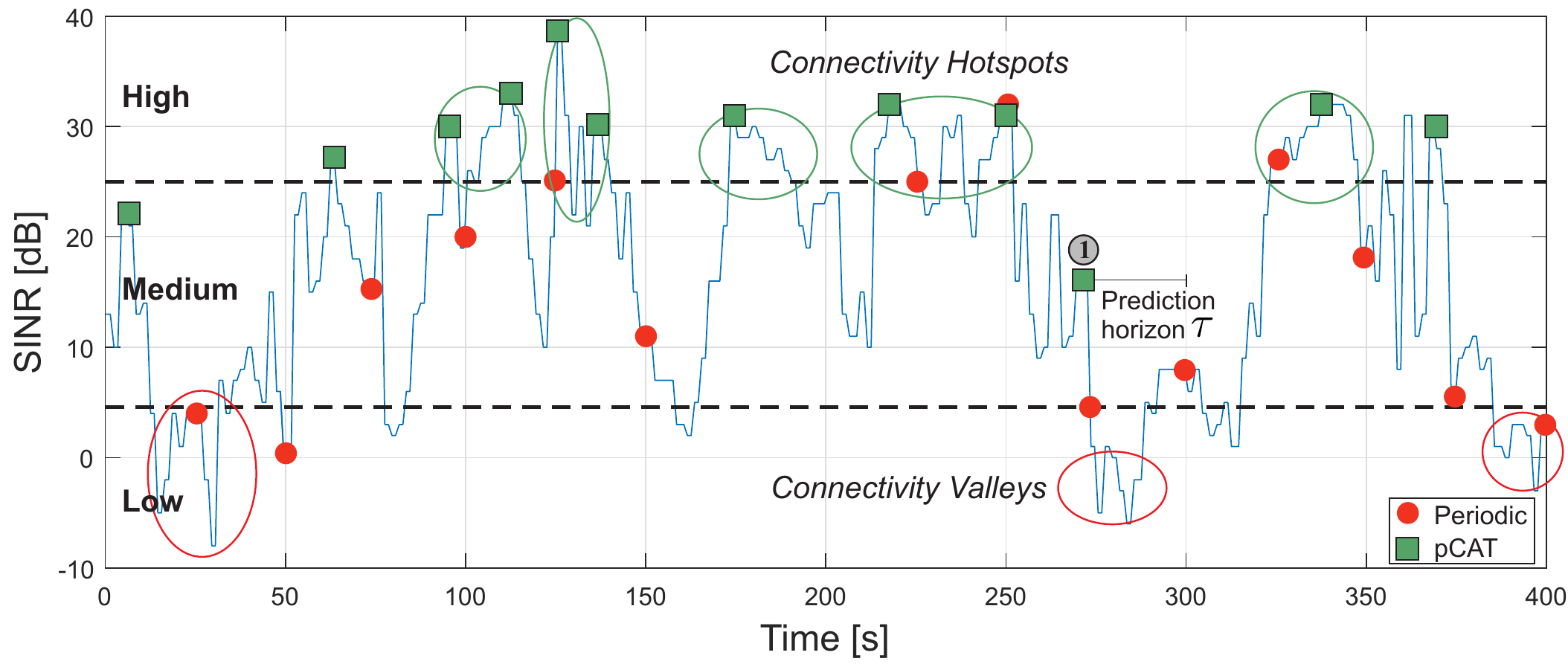}{Example temporal behavior of the proposed \ac{pCAT} transmission scheme in comparison to periodic data transmissions.}{fig:pcat_time}{0cm}{-0.7cm}{1}
An example comparison of the temporal behavior of context-predictive data transmission using \ac{SINR}-based \ac{pCAT} and straightforward fixed-interval data transfer is shown in \Fig{\ref{fig:pcat_time}}. Since the periodic approach does not consider the network quality within the transmission decision, the \ac{SINR} at the begin of the transmission is uniformly distributed over the whole value range of the \ac{SINR} and multiple transmissions are performed during low network quality periods. Contrastingly, the proposed context-predictive approach is able to avoid those resource-inefficient transmissions by exploiting \emph{connectivity hotspots}.

%
%
\subsection{Context-aware Data Transmission with \ac{CAT}} \label{sec:cat_tx}

%
%
The proposed model is based on a probabilistic process with the aim to calculate the transmission probability $p_{\Phi}(t)$ with a fixed channel assessment interval $t_{p}$ based on the measured network quality indicators. 
%
%
While the groundwork for this idea that is presented in \cite{Ide/etal/2015a} was purely focused on the \ac{SINR} for assessing the channel quality, current off-the-shelf \ac{LTE} modems and smartphones provide additional indicators, that allow a finer-grained analysis of the current connectivity situation.  
Therefore, an abstract metric $\Phi$ is introduced, which is described by its assumed minimum value $\Phi_{\text{min}}$ and its maximum value $\Phi_{\text{max}}$ that implicitly define the operation range $\Phi_{\text{max}}-\Phi_{\text{min}}$. Each indicator contained in the channel context $\vec{C}(t)$ can be mapped to a corresponding metric $\Phi_{i}(t)$.
In order to allow the comparison of multiple metrics that are related to different value ranges (e.g., \ac{RSRP} and \ac{RSRQ}), the measured metric value $\Phi(t)$ is transformed into the \emph{normed current metric value} $\Theta(t)$ with \Eq{\ref{eq:theta}}. This approach also enables the joint consideration of multiple different metrics \cite{Sliwa/etal/2018b}.
%
%
\begin{align}\label{eq:theta}
	\Theta(t) &= \frac{\Phi(t)-\Phi_{\text{min}}}{\Phi_{\text{max}}-\Phi_{\text{min}}}
\end{align}
%
%
The resulting transmission probability is then computed with Eq.~\ref{eq:cat}. With $\Delta t$ being the elapsed time since the last performed transmission, $p_{\Phi}(t)$ is computed based on the measured channel quality, if the \emph{time interval condition} $t_{\text{min}} < \Delta t < t_{\text{max}}$ is fulfilled. $t_\text{min}$ guarantees a \emph{minimum packet size} and $t_\text{max}$ specifies a \emph{maximum buffering delay} that corresponds to the actual application requirements.
%
%
\begin{align} \label{eq:cat} 
	p_{\Phi}(t) =\left\{\begin{array}{ll} 
		0 & : \Delta t \leq t_{\text{min}}\\	
		\Theta(t)^{\alpha} & : t_{\text{min}} < \Delta t < t_{\text{max}} \\ 
		1 & :  \Delta t > t_{\text{max}} \\
	\end{array}\right.
\end{align}
The formula allows to control to which extend a metric should prefer values that are close to $\Phi_{\text{max}}$ by the \emph{weighting exponent} $\alpha$. \Fig{\ref{fig:cat_analytic}} shows the resulting analytical temporal behavior for different values of $\alpha$. If the time interval condition is fulfilled and $\Phi(t)$ exceeds $\Phi_{\text{max}}$, the transmission probability is $1$ and the transmission is performed in any case.
%
%
\basicFig{}{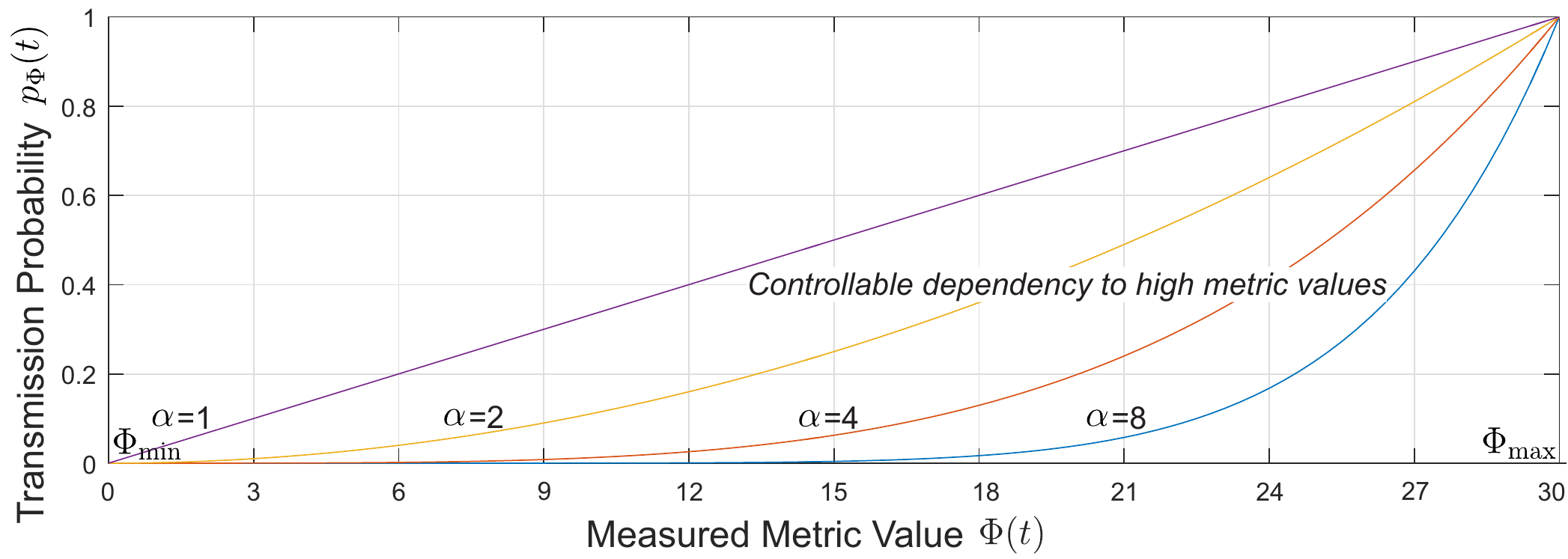}{Analytic behavior of the channel-dependent transmission probability for \ac{CAT} with different values of the weighting exponent $\alpha$.}{fig:cat_analytic}{-0.5cm}{0cm}{1}
%
%
%

%
%
%

%
%
\subsection{Context-predictive Data Transmission Exploiting Multi-layer Connectivity Maps with \ac{pCAT}} 

%
%
In the following, the previously presented \ac{CAT} scheme is extended to the context-predictive \ac{pCAT} that leverages a priori information about the channel quality along the \emph{anticipated trajectory}. The aim is to optimize the data transmission scheduling further by integrating knowledge about the future network state into the transmission process. As the basic requirement to predict the future channel context $\tilde{\vec{C}}(t+\tau)$ is the availability of a position forecast $\tilde{\vec{P}}(t+\tau)$ for a defined prediction horizon $\tau$, \ac{pCAT} is divided into a \emph{mobility prediction} component and the actual transmission process. 

%
%
\subsubsection{Mobility Prediction} \label{sec:mobility_prediction}

In the following, multiple prediction approaches, that exploit different sensors and information types and differ in the implementation complexity are discussed. Since \ac{GPS} coordinates and the \ac{WGS84} reference frame are used in the live-system, all calculations have to be performed in the orthodromic domain \cite{Weintrit/Kopacz/2012a}. Nevertheless, the formulas are presented in a cartesian coordinate system for better understanding here.
%
%
The proposed scheme focuses on the use of generic approaches that can be efficiently used in a live-system without involving a high computation overhead. More complicated approaches based on maneuver detection \cite{Houenou/etal/2013a} have been proposed in literature and will be considered for future extensions.

\paragraph{GPS-based Extrapolation}

%
%
The most straightforward approach is to extrapolate the future position by using the location, direction and velocity information provided by the \ac{GPS} receiver.
%
%
With the north-aligned angular direction $\lambda$ and the current vehicle velocity $v$, the future position is estimated with \Eq{\ref{eq:gps_prediction}}.
%
%
\begin{align} \label{eq:gps_prediction}
   \renewcommand*{\arraystretch}{1.3}
	\tilde{\vec{P}}(t+\tau) = \vec{P}(t) + 
	\begin{pmatrix}
		{\sin(\frac{\Pi}{2})} \cdot \cos\left( \frac{\lambda \cdot \Pi }{180^{\circ}} \right) \\
		{\sin(\frac{\Pi}{2})} \cdot  \sin\left( \frac{\lambda \cdot \Pi }{180^{\circ}} \right)
	\end{pmatrix} 	
	\cdot \tau
	\cdot v
\end{align}
%
%
The advantage of using extrapolation is that it can be implemented in a very simple manner by using only the currently measured \ac{GPS} information. However, it assumes the direction $\lambda$ and the velocity $v$ to be constant for the duration of $\tau$. Therefore, the resulting prediction accuracy is highly reduced --- especially for larger values of $\tau$ --- if the vehicle turns, encounters stop-and-go traffic or is influenced by traffic signals and other traffic participants.

\paragraph{Leveraging Trajectory-knowledge from the Navigation System} \label{sec:trajectory_prediction}

%
%
For overcoming the limitations of the previous approach, mobility prediction based on \emph{trajectory} information is now discussed. While \emph{planned trajectories} might be accessible through a direct interface to the navigation system --- which will likely be the case for upcoming automated vehicles --- this type of information could also be derived by exploiting the regularities in human behavior itself. In fact, the analysis in \cite{Song/etal/2010a} points out that 95\% of human mobility can be predicted by exploiting people's regular movement on the same paths (e.g., the way to work or to grocery stores).
%
%
\\
With the assumption of having data for the same track from multiple trips available, the segment-wise mean trajectory is calculated in a preprocessing step with the approach presented in \cite{Niehoefer/etal/2009a}.
%
%
During the online mobility prediction, the current trip is detected by the highest matching of the measured position and direction values to all locally stored trips. 
$\tilde{\vec{P}}(t+\tau)$ is then derived by virtually moving the vehicle along the anticipated path for a duration of $\tau$ in an iterative process. For each prediction, the movement potential $\tilde{D}=v\cdot\tau$ is computed and the traveled distance $D$ is initialized with $D=0$. In each iteration $i$, $D$ is incremented by the distance  $d_{i,j}=||\vec{W}_{j}-\vec{W}_{i}||$ between the consecutive waypoints $\vec{W}_{i}$ and $\vec{W}_{j=i+1}$. When $D$ exceeds $\tilde{D}$, the final position is obtained from interpolation using \Eq{\ref{eq:interpolation}}.
%
%
\begin{align} \label{eq:interpolation}
	\tilde{\vec{P}}(t+\tau) = \vec{W}_{i}  
	+ \frac{\vec{W}_{j}-\vec{W}_{i}}{||\vec{W}_{j}-\vec{W}_{i}||} 
	\cdot \left( D - \tilde{D} - d_{i,j}\right) 
\end{align}

\paragraph{Lightweight Trajectory-aware Approach - Prediction based on a Reference Trace}

%
%
A lightweight alternative, that requires less data than the previous approach  is to utilize the last measurements of the same track as a \emph{reference trace}.
The mobility prediction process itself is then equal to the one presented in \Sec{\ref{sec:trajectory_prediction}}, but the preprocessing stage is omitted as only a single track is utilized for the computation of the future position. Analogously, the connectivity map only contains the values of a single measurement drive per distinct track.
The price to pay for the increased resource efficiency is the loss of the \emph{aggregation gain} that is obtained by cell-wise averaging, which reduces the impact of outlier measurements, especially for highly dynamic metrics as the \ac{SINR}. In \Sec{\ref{sec:mobility_prediction_accuracy}}, the resulting error for the network quality prediction is discussed for the considered mobility prediction methods.

\subsubsection{Context-predictive Transmission Process} \label{sec:pcat_tx}

With the predicted position $\tilde{\vec{P}}(t+\tau)$ being available after the mobility prediction step, the future channel context $\tilde{\vec{C}}(t+\tau)$ is looked up from the connectivity map as illustrated in \Fig{\ref{fig:connectivity_map}}. The cell index $m$ for the corresponding entry is obtained with \Eq{\ref{eq:cellIndex}} for a defined cell size $c$.
%
%
\begin{align}\label{eq:cellIndex}
	m = \left\lfloor \frac{\tilde{\vec{P}}(t+\tau)}{c}\right\rfloor
\end{align}
%
%
\basicFig{}{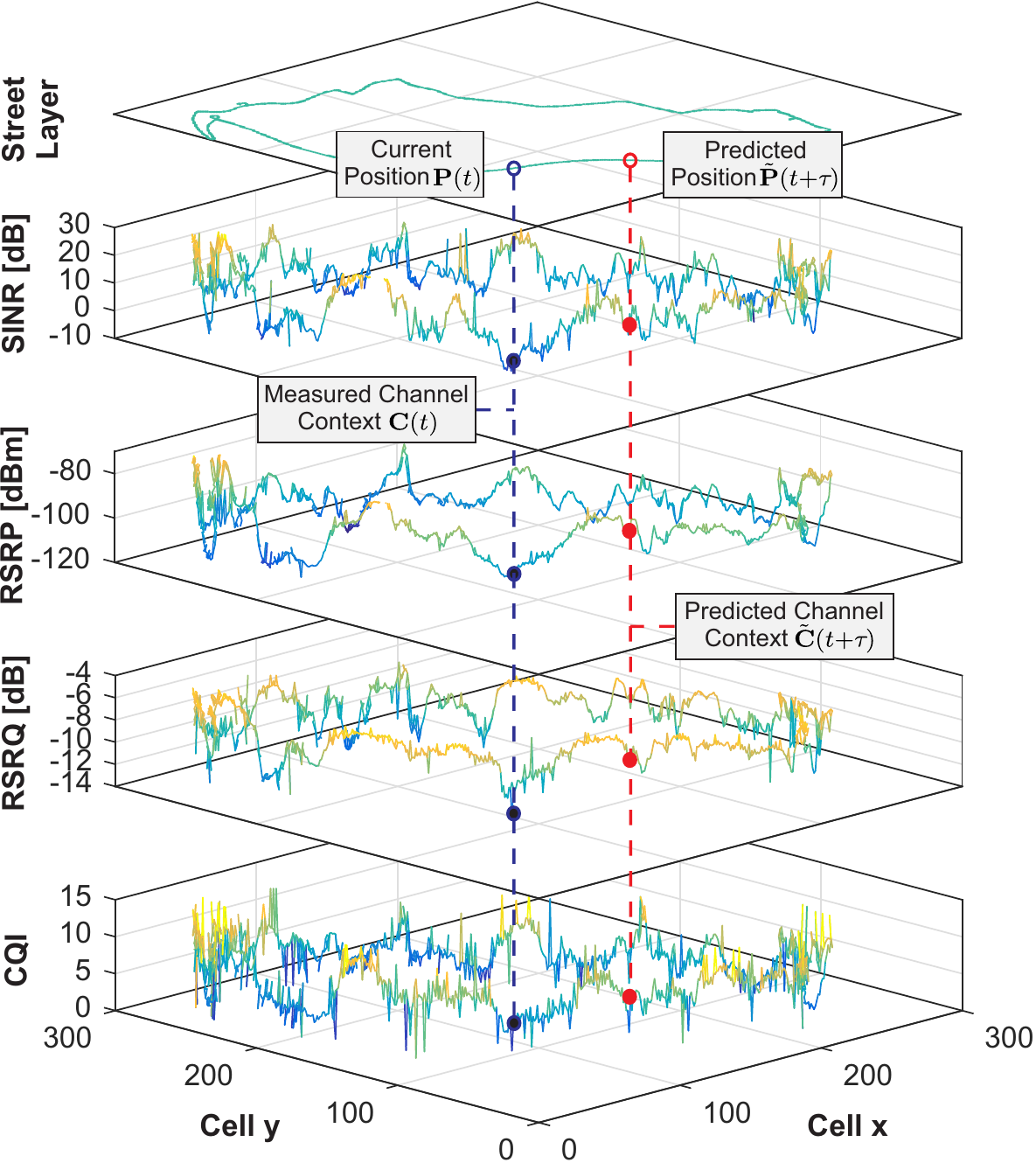}{Multi-layer connectivity maps as an enabler for anticipatory communication. With the help of mobility prediction, the current measured channel quality can be compared to its predicted future state.}{fig:connectivity_map}{-0.5cm}{0cm}{1}
The predicted metric value $\tilde{\Phi}(t + \tau)$ is extracted from $\tilde{\vec{C}}(t+\tau)$ and the anticipated gain $\Delta \Phi(t)$ is computed using \Eq{\ref{eq:delta_phi}}.
%
%
\begin{align}\label{eq:delta_phi}
	\Delta \Phi(t) &= \tilde{\Phi}(t + \tau) - \Phi(t)
\end{align}
Analogously to \ac{CAT}, the transmission probability $p_{\Phi}(t)$ is then computed with respect to the defined timeouts using \Eq{\ref{eq:pCAT}}. For the consideration of the channel quality development, the \ac{pCAT}-specific exponent $\beta$ is introduced, which controls the impact of the context prediction within \Eq{\ref{eq:pCat_exponent}}. 
%
%
\begin{align} \label{eq:pCAT} 
	p_{\Phi}(t) =\left\{\begin{array}{ll} 
		0 & : \Delta t \leq t_{\text{min}}\\	
		\Theta(t)^{\alpha \cdot z} & : t_{\text{min}} < \Delta t < t_{\text{max}} \\ 
		1 & :  \Delta t > t_{\text{max}} \\
	\end{array}\right.
\end{align}
%
%
\begin{align} \label{eq:pCat_exponent} 
   \renewcommand*{\arraystretch}{1.2}
	z =\left\{\begin{array}{ll} 
		\max\left( \left| \Delta  \Phi(t) \cdot \left( 1-\Theta(t) \right) \cdot \beta \right| , 1 \right) & : \Delta \Phi(t) > 0\\	
		\left( \max\left( \left| \Delta  \Phi(t) \cdot \Theta(t)\cdot \beta \right| , 1 \right) \right)^{-1} & : \Delta \Phi(t) \leq 0 \\ 
	\end{array}\right.
\end{align}
%
%
\basicFig{b}{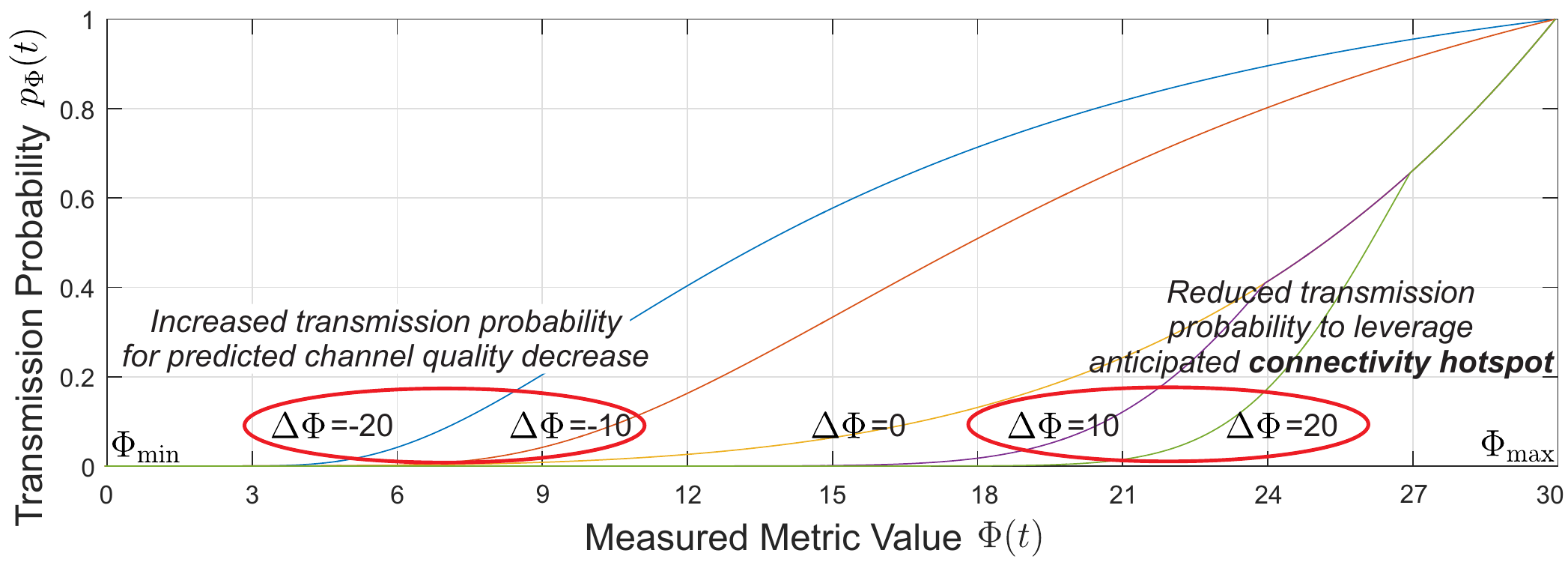}{Analytical behavior of the resulting transmission probability for \ac{pCAT} ($\alpha=4$) with respect to the current metric value $\Phi(t)$ for different values of the expected channel development $\Delta\Phi(t) $.}{fig:pcat_analytic}{0cm}{-0.5cm}{1}
%
%
\Fig{\ref{fig:pcat_analytic}} shows the resulting analytical behavior of the transmission probability for different values of $\Delta\Phi$. While the scheme behaves equal to \ac{CAT}  for $\Delta\Phi=0$, \ac{pCAT} sends earlier if it expects the channel quality to decrease in the future ($\Delta\Phi<0$) and schedules the transmission to a later point of time if it anticipates an improvement for the network quality ($\Delta\Phi>0$).  
\\
%
%
Although \ac{pCAT} only considers two discrete points of time $t$ and $t+\tau$ for its decision, it in fact behaves like a moving window as the vehicle moves forward on the track and the transmission decision is calculated frequently with respect to the channel assessment interval $t_{\text{p}}$.
%
%
If any of the prediction steps fails (e.g., due to missing \ac{GPS} signal or if the map does not contain data for the predicted cell), \ac{pCAT} performs a \emph{context-aware fallback} by switching to the purely probabilistic \ac{CAT} model.
%
%
Although the necessary data collection highly benefits from the mutual synergies of a crowdsensing-based approach, the whole proposed scheme can also be implemented in a local sandbox for dealing with any privacy-related concerns.

%
%
%

%
%
\subsection{Machine Learning-based Data Rate Prediction} \label{sec:data_rate_prediction}

Predicting the data rate is a supervised learning task. Given the
features $\vec{X}$ of the measurements, a prediction model $M$ that
predicts the data rate $S$ is applied. Different model classes $M$ are possible, each
representing a different function class $f:\vec{X} \rightarrow S$ and pertaining a
different list of parameters $\vecg{\psi}$. Due to the supervised
learning task, $\operatorname{arg\,min}_{f_{\vecg{\psi}} \in M} l(f_{\vecg{\psi}}(\vec{X}), S)$, labeled data is available as ground truth for the data rate. The error of a particular model function $f$ may thus be assessed by comparison of
the predicted data rate to the known data rate. Possible loss functions $l$ are the
\ac{MAE}, which is a measure for the absolute distance among prediction
and the true label, and the \ac{RMSE} measuring the Euclidean error.
\begin{align}
\operatorname{MAE}(f(\vec{X}), S)=\frac{1}{N}\sum_{i=1}^N\left|S^{(i)}-f(\vec{X}^{(i)})\right|\\
\operatorname{RMSE}(f(\vec{X}), S) = \sqrt{\frac{1}{N} \sum_{i=1}^N \left(S^{(i)}-f(\vec{X}^{(i)}) \right)^2}
\end{align}

Several model classes are possible. In order to decide on a model class, not only the
performance on training instances is important but also the validation of
novel, yet unseen, examples. Otherwise it could possibly happen that a model
learns all examples `by heart' and has perfect prediction performance on
training instances but does not generalize at
all - this phenomenon is defined as \emph{overfitting}. Possible countermeasures against
overfitting exist, e.g., recording a larger more diverse
data set or choosing a model with less capacity, or including the capacity of the
model in the objective function via a regularization term. The details on
regularization are beyond the scope of this paper and can be found in \cite{trevor2009elements}. The  model selection is performed with the tool WEKA \cite{Hall/etal/2009a} and different model classes are tested. In this paper, the applied models are regression tree (in particular, the \ac{M5T}~\cite{Quinlan/1992a}), \ac{RF}~\cite{Breiman/2001a}, \ac{LR}, \ac{ANN}, and \ac{SVM}
\cite{scholkopf2001learning}. All methods are discussed in
\cite{trevor2009elements}, thus a brief overview of selected methods is provided here.

The simplest model is \ac{LR} which fits a linear combination
of the input features $\vec{X}$ to the output label $S$. 
A more sophisticated way is the split of the data set into regions and
application of different linear models within different regions. The regression
tree performs exactly these splits using axis parallel hyperplanes by comparison
of each feature with a threshold. This distinction of feature vectors
based on thresholds of its features is captured in a tree structure. In its
leafs, different linear models are applied. The \ac{RF} trains not one
regression tree but multiple and learns a linear combination of their
predictions. The theoretical aspects of artificial neural networks and support vector
machines are out of scope of this paper, additional information can be found in
\cite{trevor2009elements}.
%
%
\renewcommand{\entry}[4]{#1 & #2 & #3 & #4 \\}%
\renewcommand{\head}[4]{\toprule \entry{\textbf{#1}}{\textbf{#2}}{\textbf{#3}}{\textbf{#4}}\midrule}%
\renewcommand{\baselinestretch}{1}%
\begin{table}[ht]
	\centering

	\caption{Performance of regression models: Artificial Neural Network (ANN), Linear Regression (LR), M5 decision Tree (M5T), Random Forest (RF) and Support Vector Machine (SVM) are compared in terms of correlation, mean
		absolute error (MAE) and root mean squared error (RMSE).}
	\vspace{-0.2cm}
	\begin{tabular}{llll}
		
		\head{Model}{Correlation}{MAE [MBit/s]}{RMSE [MBit/s]}
		
		\entry{ANN}{0.83}{1.62}{2.07}
		\entry{LR}{0.57}{2.19}{2.92}
		\entry{\textbf{M5T}}{\textbf{0.86}}{\textbf{1.33}}{\textbf{1.81}}
		\entry{\textbf{RF}}{\textbf{0.86}}{\textbf{1.3}}{\textbf{1.79}}
		\entry{SVM}{0.80}{1.161}{2.13}

		\bottomrule
		
	\end{tabular}
	\label{tab:ml-results}
	\vspace{-0.3cm}
\end{table}%
\renewcommand{\baselinestretch}{\BLS}
%
%
The evaluation of the prediction methods is performed based on more than 2500 real-world measurements of periodic and \ac{CAT}-based transmissions that were performed in the context of earlier work in \cite{Sliwa/etal/2018b} on two different tracks (details about the measurement setup are provided in \Sec{\ref{sec:methodoloy}}). The feature set is formed by \ac{RSRP}, \ac{RSRQ}, \ac{SINR}, \ac{CQI} and velocity measurements in combination with the payload size of the data packets. The label is defined as the measured data rate of the active transmissions.
\Tab{\ref{tab:ml-results}} shows the resulting prediction performance for the considered models and evaluation metrics. Although the absolute highest accuracy is achieved with the \ac{RF} model, it only performs slightly better than the \ac{M5T} approach. 
\\
%
%
\renewcommand{\entry}[5]{#1 & #2 & #3 & #4 & #5 \\}
\renewcommand{\head}[5]{\toprule \entry{\textbf{#1}}{\textbf{#2}}{\textbf{#3}}{\textbf{#4}}{\textbf{#5}}\midrule}
\renewcommand{\baselinestretch}{1}
\begin{table}[ht]
	\centering

	\caption{Correlation results for data rate prediction of machine learning models divided by suburban track (S) and highway track (H)}
	\vspace{-0.2cm}
	\begin{tabular}{lllll}
		
		\head{Model}{S}{S$\to$H}{H}{H$\to$S}
		
		\entry{ANN}{0.842}{0.839}{0.829}{0.819} 
		\entry{LR}{0.511}{0.727}{0.725}{0.514}  
		\entry{M5T}{\textbf{0.855}}{\textbf{0.829}}{\textbf{0.858}}{\textbf{0.798}} 
		\entry{RF}{\textbf{0.872}}{\textbf{0.844}}{\textbf{0.868}}{\textbf{0.862}} 
		\entry{SVM}{0.799}{0.572}{0.813}{0.699} 

		\bottomrule
		
	\end{tabular}
	\label{tab:learningTracks}
	\vspace{-0.2cm}
\end{table}
\renewcommand{\baselinestretch}{\BLS}
In order to evaluate the generalizability of the learned relationship between context parameters and data rate which mirrors the dependency of the models of the tracks themselves, a cross-check with separated data sets --- suburban (\textbf{S}) and highway (\textbf{H}) --- is performed. \Tab{\ref{tab:learningTracks}} shows the resulting correlation between predicted and measured data rate for the different models. \textbf{S} and \textbf{H} show the 10-fold cross validation for each of the tracks. For \textbf{S}$\to$\textbf{H} and \textbf{H}$\to$\textbf{S}, the trained model is tested on the measurement data of the other track.
It can be seen that the behaviors differ significantly for the different models. While \ac{ANN}, \ac{M5T} and \ac{RF} show a similar behavior for all of the variants, the \ac{SVM} accuracy is significantly reduced for \textbf{S}$\to$\textbf{H} and \textbf{H}$\to$\textbf{S}. The \ac{LR} model does not work well on the suburban data set.

%
%
Based on the prediction results, the \ac{M5T} model is chosen as the applied prediction model within the application phase. It achieves a good overall performance, allows a very \emph{lightweight implementation} and can be used for online data rate prediction without causing significant computation overhead. For simplicity, in the following, the usage of the $\Phi_{\text{M5T}}$ metric for sensor data transmissions will be referred to as \textbf{\ac{ML-CAT}}, respectively \textbf{\ac{ML-pCAT}}.

%
%
\basicFig{}{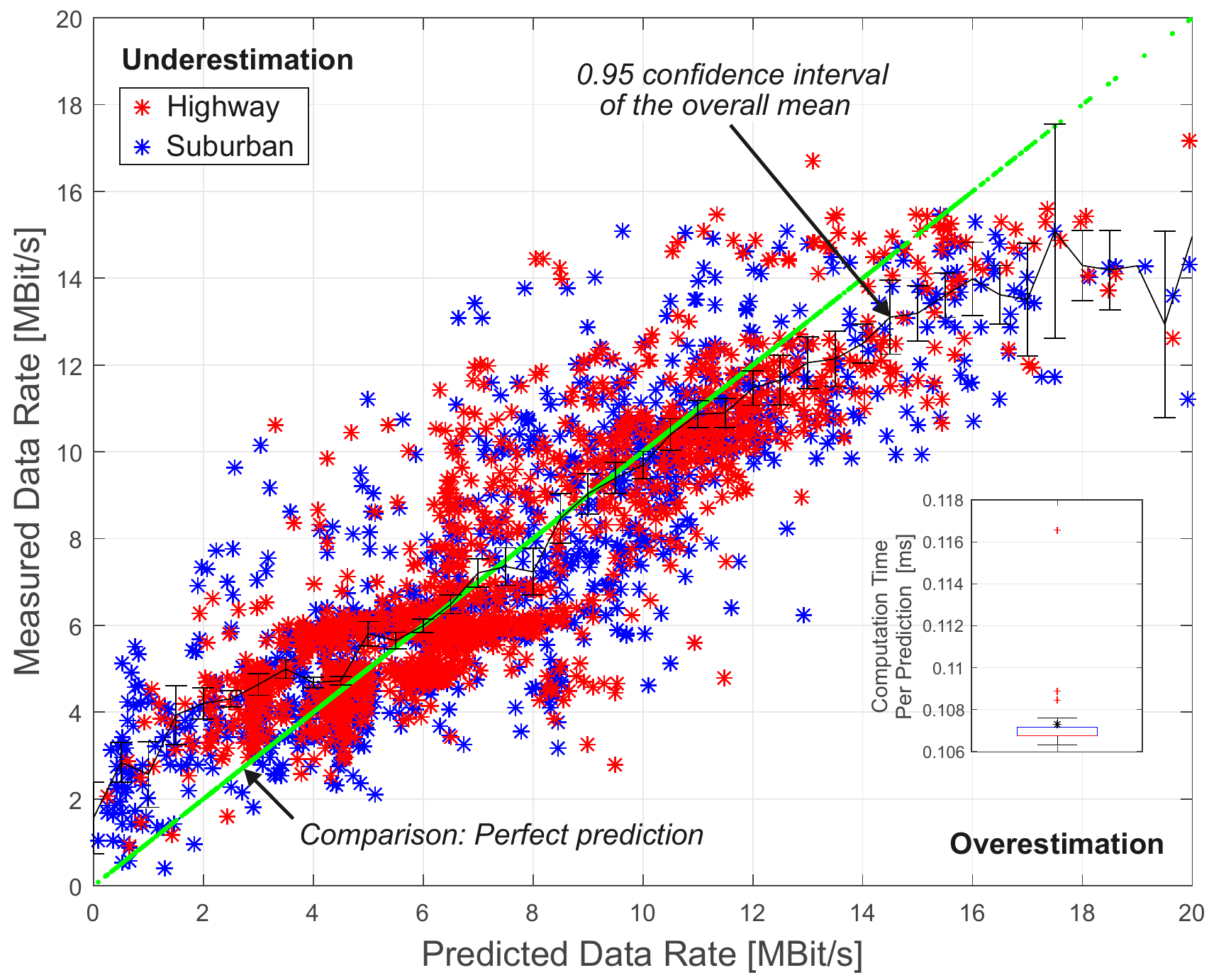}{Achieved accuracy of the \ac{M5T} data rate prediction scheme. Blue markers illustrate transmissions on the suburban track, red markers show the performance on the highway track. The overlay shows the measured computation time per prediction on the Android device.}{fig:m5t_eval}{-0.5cm}{0cm}{1}
\Fig{\ref{fig:m5t_eval}} shows the resulting prediction accuracy using \ac{M5T}. The left upper triangle shows the \emph{underestimation} area and the lower right triangle represents the \emph{overestimation} area. From the application-centric perspective, underestimations are not considered harmful, as the transmission even achieves a higher data rate than expected.

\section{Methodological Setup of the Empirical Performance Evaluation} \label{sec:methodoloy}

In this section, the methodological aspects of the real world evaluation and the considered \acp{KPI} for the performance evaluation are introduced.

\subsection{Real World Evaluation Scenario} 

In order to evaluate the properties of the proposed transmission schemes in a realistic scenario, a comprehensive empirical performance evaluation is performed in the public cellular \ac{LTE} network and within a vehicular context. \Tab{\ref{tab:parameters}} provides a summary of the application-related key parameters.
\renewcommand{\entry}[2]{#1 & #2 \\}
\renewcommand{\head}[2]{\toprule \entry{\textbf{#1}}{\textbf{#2}}\midrule}
\renewcommand{\baselinestretch}{1}
\begin{table}[ht]
	\centering
	\caption{Parameters of the Reference Scenario}
	\vspace{-0.2cm}
	\begin{tabular}{ll}
		
		\head{Parameter}{Value}
		
		\entry{Sensor frequency $f_{\text{sensor}}$}{1~Hz}
		\entry{Sensor payload size $s_{\text{sensor}}$}{50~kByte }
		\entry{Channel assessment interval $t_{\text{p}}$}{1~s}
		\entry{Minimum delay between two transmissions $t_{\text{min}}$}{10~s}
		\entry{Maximum buffering time $t_{\text{max}}$}{120~s}
		\entry{Prediction horizon $\tau$}{$\left\lbrace 10, 30, 60 \right\rbrace$~s}
		\entry{Connectivity map cell width $c$}{25~m}
		
		\bottomrule
		
	\end{tabular}
	\label{tab:parameters}
	\vspace{-0.2cm}
\end{table}
\renewcommand{\baselinestretch}{\BLS}%
Channel sensing and data transmission are handled by an Android-based application (executed on a Samsung Galaxy S5 Neo - Model SM-G903F), which is provided in an Open Source way\githubUrl. Sensor packets of size $s_{\text{sensor}}$ are generated by a virtual sensor application with a sensor frequency $f_{\text{sensor}}$ and stored in a local buffer until a transmission decision is made for the whole data buffer. For each $t_{\text{p}}$, the channel is sensed and the transmission probability is computed. All data transmissions are performed in the \ac{LTE} uplink to a cloud server using \ac{HTTP} POST.
%
%
The different considered metrics and their respective parameterizations are shown in \Tab{\ref{tab:metrics}}. \ac{CAT} with the $\Phi_{\text{SINR}}$ metric is equal to the legacy version presented in \cite{Ide/etal/2015a}, analogously, \ac{pCAT} with $\Phi_{\text{SINR}}$ has the same behavior as \cite{Wietfeld/etal/2014a}. The values for the weighting factor $\beta$ have to be chosen with respect to the metric's value range and its granularity. In order to allow a fair comparison among the different metrics, $\beta$ is chosen with respect to the definition of $\Phi_{\text{SINR}}$ metric and its respective value range with \Eq{\ref{eq:beta}}.
\renewcommand{\entry}[6]{#1 & #2 & #3 & #4 & #5 & #6\\}
\renewcommand{\head}[6]{\toprule \entry{\textbf{#1}}{\textbf{#2}}{\textbf{#3}}{\textbf{#4}}{\textbf{#5}}{\textbf{#6}} \midrule}
\renewcommand{\baselinestretch}{1}
\begin{table}[ht]
	\centering
	\caption{Parameters for the considered metrics}
	\vs{-0.7cm}
	\begin{tabular}{llllll}
		
		\head{}{$\Phi_{\text{RSRP}}$}{$\Phi_{\text{RSRQ}}$}{$\Phi_{\text{SINR}}$}{$\Phi_{\text{CQI}}$}{$\Phi_{\text{M5T}}$}
		
		\entry{min}{-120 dBm}{-11 dB}{0 dB}{2}{0 MBit/s}
		\entry{max}{-70 dBm}{-4 dB}{30 dB}{16}{18 MBit/s}
		\entry{$\alpha$}{8}{8}{8}{8}{8}
		\entry{$\beta$}{0.3}{2.14}{0.5}{1.07}{1}

		\bottomrule
		
	\end{tabular}
	\label{tab:metrics}
	\vs{-0.7cm}
\end{table}
\renewcommand{\baselinestretch}{\BLS}

%
%
\begin{equation} \label{eq:beta}
	\Phi_{i,\beta} = \Phi_{\text{SINR},\beta} \cdot \frac{\Phi_{\text{SINR,max}}-\Phi_{\text{SINR,min}}}{\Phi_{\text{i,max}}-\Phi_{\text{i,min}}}
\end{equation}
%
%
For the context-predictive \ac{pCAT}, the connectivity map and the trajectory prediction utilize the acquired data of the \ac{CAT} evaluation phase.

%
%
The raw data of the experimental evaluation of the different transmission schemes is provided at \cite{Sliwa/2018a} and the measurement software as well as the obtained data samples for the transmission power estimation can be accessed via \cite{Falkenberg/2018a}.

%
%
\basicFig{}{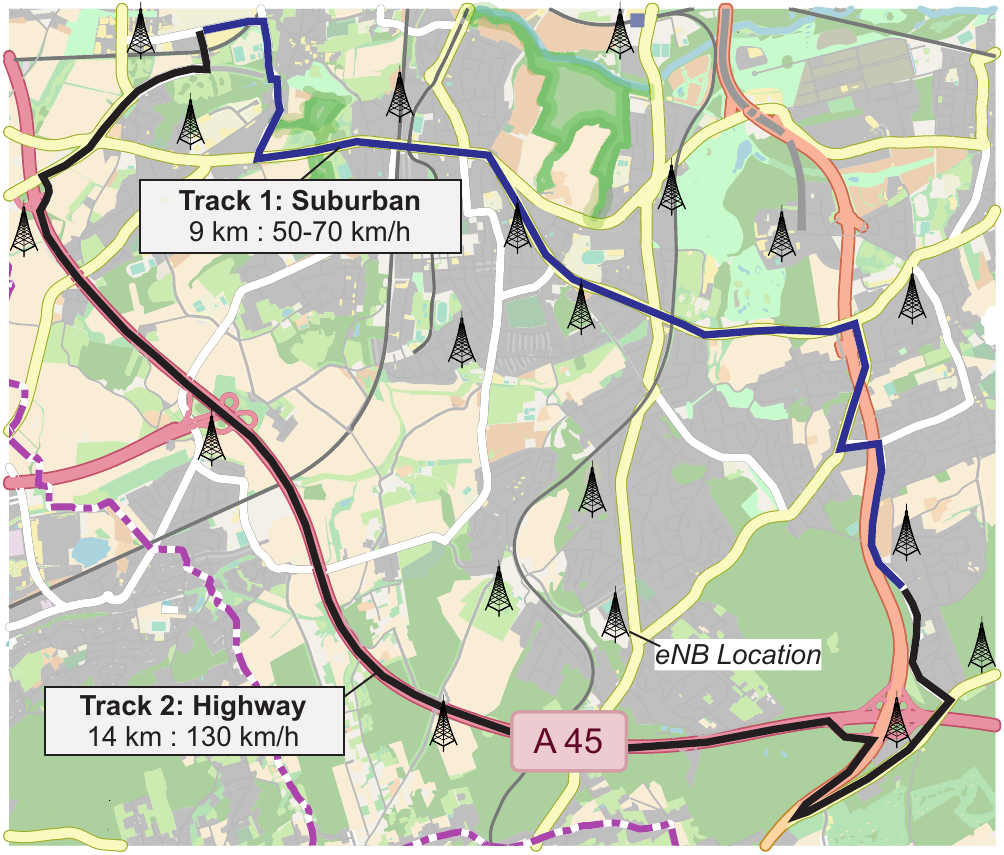}{Street map and network infrastructure of the evaluation scenario consisting of suburban and highway tracks. (Map: ©OpenStreetMap contributors, CC BY-SA.)}{fig:map}{-0.5cm}{0cm}{1}
Fig.~\ref{fig:map} shows the street map with the different tracks used for the experimental performance evaluation. 
\begin{itemize}
	\item \textbf{Track~1:} Suburban roads with upper speed limits in the range of 50-70~\si{\kilo\meter\per\hour} (\SI{14}{\kilo\meter})
	\item \textbf{Track~2:} Highway traffic with upper speed limits of \SI{130}{\kilo\meter\per\hour} (\SI{9}{\kilo\meter})
\end{itemize}
Each parameterization of the transmission schemes has been evaluated five times on each of the tracks. Overall, more than 7500 data transmissions were performed within a total driven distance of more than \SI{2000}{\kilo\meter}. On the application layer, all data transmissions were performed successfully.

\subsection{Key Performance Indicators} 

In the result section of this paper, the performance of different transmission schemes is compared in three dimensions.

%
%
%
	\textbf{End-to-end Data Rate:} The data rate evaluations are performed on the application level and are considered as a measurement for the transmission-efficiency of data packets. Moreover, since high data rates indicate short transmission durations, data rate optimization is also related to early release of occupied spectrum resources. The potentials for improving the coexistence of different resource-consuming cell users are further investigated in \cite{Ide/etal/2015a}. 
	
	\textbf{\acf{AoI}:} \ac{AoI} is a metric for the freshness of information of delay-tolerant applications such as crowdsensing and data analytics. Therefore, it provides a better match with the considered crowdsensing use-case than delay measurements. In the considered definition $\text{AoI} = t_\text{app} - t_\text{gen}$, it covers the time from the data generation $t_\text{gen}$ (e.g., the actual measurement report of a physical sensor) to the reception time $t_\text{app}$ of the information by the processing application and also includes the delays caused by buffering and transmission. From an application point of view, information is considered useless if a certain \ac{AoI} value is exceeded. For the proposed transmission scheme, the resulting \ac{AoI} is mainly impacted by the buffering delay and can be controlled with the timeouts $t_{\min}$ and $t_{\max}$. Since all transmitted data packets contain multiple measurement values that have individual times of generation, the \ac{AoI} of a data packet is defined as the mean age of all contained sensor measurements.

	\textbf{Energy-efficiency:} 
	Although the real world evaluations are carried out with a ground-based vehicular setup, where the energy consumption of the communication system is negligible, the achieved results and methods are relevant to related research fields. The integration of small-scale \acp{UAV} into upcoming \acp{ITS} -- e.g., for near field delivery and aerial sensing -- is currently receiving great attention within the scientific community \cite{Menouar/etal/2017a}. As these vehicles are severely constrained by the available energy resources, the improvement of the overall power consumption is of great importance for optimizing the total operation time \cite{Zeng/Zhang/2017a}. In the context of \ac{IoT}-enabled logistics \cite{Zanella/etal/2014a}, smart containers, which are equipped with battery powered communication systems, report status information and sensor measurements (e.g., temperature values) to enable continuous tracking and monitoring. In addition, autonomous systems within the car itself are powered by dedicated batteries (e.g., anti-theft systems). The results in Sec.~\ref{sec:results} show that the proposed transmission scheme has a severe impact on the overall power consumption of the mobile device. Therefore, we consider the energy-efficiency as one of the \acp{KPI}, as it allows us to point out improvement potentials for the related application fields, which typically implement periodic data transmission mechanisms.

	The uplink power consumption $P_\text{UL}$ is mainly depending on the actual transmission power $P_\text{TX}$ of the \ac{UE} and related to the different amplification states of the power amplifiers \cite{Jensen/etal/2012a}. Since $P_\text{TX}$ is usually not reported by embedded devices and smartphones, a mechanism for accessing this hidden parameter using machine learning and direct modem interfacing is presented in \Sec{\ref{sec:tx_power}}. 


\section{Machine Learning-enabled Post-processing Uplink Power Consumption Analysis} \label{sec:tx_power} \label{sec:copomo}

\begin{figure*}[] 
	\centering
	\includegraphics[width=1\textwidth]{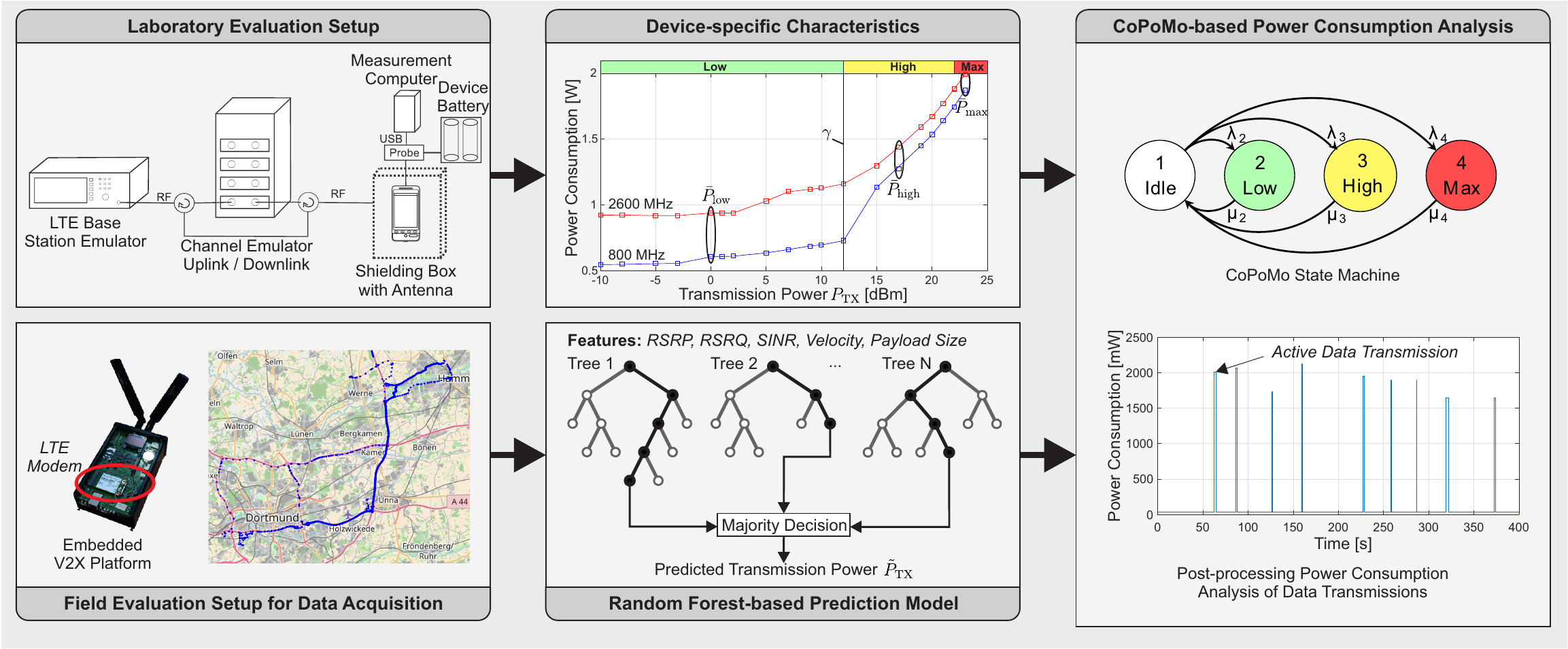}
	\vspace{-0.5cm}
	\caption{Overall architecture model for the closed process for power consumption analysis with missing information. The device-specific characteristics are captured by laboratory measurements and the relationship between transmission power and channel indicators is learned using the field evaluation setup of \cite{Falkenberg/etal/2018a} (Map: ©OpenStreetMap contributors, CC BY-SA.).}
	\label{fig:copomo}
	\vspace{-0.5cm}
\end{figure*}

%
%
Performing communication-related power consumption measurements of an embedded device is a non-trivial task that requires precise isolation of physical components and involved software. As numerous components of the system, such as modem and processor, are highly integrated into a single \SoC, it is not possible to measure the modem's power consumption isolatedly. On the other hand, measurements of the whole system's power consumption are superimposed by any other system activities, \eg \GPS, IO operations, and background services. Consequently, such measurements are rather performed within a laboratory setup and not with a highly integrated mobile system. 
In order to allow the evaluation of the energy-efficiency of \ac{CAT} on (mobile) embedded platforms, a closed process for isolated uplink power consumption analysis is derived in the following that combines existing models and approaches \cite{Dusza/etal/2013a, Falkenberg/etal/2017a, Falkenberg/etal/2018a}. \Fig{\ref{fig:copomo}} shows the overall architecture model. Accurate laboratory measurements are used to obtain the device-specific behavior characteristics of power consumption with regard to the applied transmission power.
%
%

%
%
The \ac{CoPoMo}~\cite{Dusza/etal/2013a} is a validated model, which uses device-specific characteristics obtained from laboratory evaluations to estimate the power consumption for uplink transmissions as a function of the transmission power within a state machine. \Fig{\ref{fig:copomo}} includes the characteristic curve of the Galaxy S5 Neo smartphone operating in two different frequency bands for a transmission power range of \SI{-10}{\dBm} to \SI{23}{\dBm}~\cite{Falkenberg/etal/2017a}. For a single frequency band, \eg, the blue curve, the characteristics can be approximated by two linear functions of different slope, which are separated by a device-specific break point $\gamma$. This behavior is caused by switching between two different internal power amplifiers. \ac{CoPoMo} further reduces the characteristics to a probabilistic four-state power model with negligible loss of accuracy. Depending on the radio conditions, a \UE uploads its data with \Low, \High, or \Max power and enters \Idle mode afterwards. The state transitions are modeled by the transition probabilities $\lambda$ and $\mu$ that are obtained from the corresponding data set. Calculating the equilibrium state of the Markovian chain finally provides an estimate for the \UE power consumption in the given scenario.

%
%
Unfortunately, in the context of this paper, the model cannot directly be applied as embedded operating systems (e.g., Android) do not provide information about the currently used transmission power and therefore circumvent the determination of the current power state.
As a consequence, this paper applies a novel machine learning-based approach for power estimation \cite{Falkenberg/etal/2018a}, which is based on the available \ac{LTE} downlink indicators.
According to the \LTE standard \cite{3GPP/2016a}, \UEs choose their transmission power $P_{\text{TX}}$ based on \Eq{\ref{eq:tx_power_standard}}:
%
%
\begin{align} \label{eq:tx_power_standard}
P_\text{TX} = \min\left(\begin{array}{l}
P_\text{max},\\
P_{0} + 10 \log_{10}(M) + \alpha \cdot PL + \Delta_\text{MCS} + \delta
\end{array}
\right).
\end{align}
$P_0$ is broadcasted by the \eNB and depicts the target \SINR per \PRB of the received signal at the \eNB.
Thus, the \UE at least has to compensate the estimated path loss $PL$, which is derived from \RSRP and the actual transmission power of the \eNB. It is weighed by a \FPC $\alpha$, which is also configured by the base station.
An additional offset $\Delta_\text{MCS}$ ensures a sufficient \SINR for the selected \MCS.
Finally, the transmission power needs to be increased according to the number of emitted \PRBs $M$ in order to keep the received \SINR constant regardless of the number of allocated resources.

The closed-loop component $\delta$ is an absolute or accumulated offset, which is transmitted in \TPC commands by the base station in \PDCCH together with the resource allocations. 
By this approach, the \eNB can increase or turn down the output power of the \UE in a feedback loop.
However, the \eNB never transmits an absolute $P_{\text{TX}}$ value to the \UE.

Since $P_\text{max}$, $P_{0}$, and $\alpha$ can be seen as network constants and $\delta$ should average to $0$ for a well-configured network, only $M$, $PL$, and $\Delta_\text{MCS}$ have to be obtained in order to estimate the transmission power $\tilde{P}_{\text{TX}}$ at application layer.
Although these remaining variables are not accessible as well, they are tightly related to observable indicators. The path loss $PL$ is internally calculated from the \RSRP.	However, the \eNB's reference signal transmit power still remains unknown to the application layer. The number of allocated \PRBs $M$ corresponds to the resulting data rate at a given \MCS. Unfortunately, in case of a prediction, the data rate is not available as an indicator. However, assuming a non-congested network, the data rate follows the \TCP slow start mechanism during the transmission, which depends on the upload size and the maximum achievable rate. The latter is related to $PL$, since $M$ and $\Delta_\text{MCS}$ are capped by $P_\text{max}$ for large $PL$. For $\Delta_\text{MCS}$, the lookup table provides \MCS-dependent power offsets to ensure a proper \SINR for a correct demodulation and decoding at the base station.	Interference and mobility (fast fading) adversely affect the \MCS, which can be indicated by \RSRQ and the \UE's velocity.	According to the \UE's power headroom, the \eNB may select the highest possible \MCS to maximize throughput and spectral efficiency. Hence, this value is also related to $PL$ and $M$.

%
%
However, the exact relationship of these variables is blurred by case differentiation and operator-specific configurations of their base stations, which makes analytical approaches complex and impractical.
Therefore, a data-driven approach leverages machine learning to obtain a prediction model for $\tilde{P}_{\text{TX}}$, which is presented in~\cite{Falkenberg/etal/2018a}.
The work analyzes the usage of different indicators and machine learning techniques for the estimation of $\tilde{P}_{\text{TX}}$ in detail.
It also provides different prediction models for \emph{simulations}, \emph{practical} applications and \emph{detailed analysis}, which differ in the number of available indicators for this task.
The data is obtained in excessive field measurements of drive tests in public cellular networks.
An overview of the covered trajectory is shown in \Fig{\ref{fig:copomo}} and covers urban, suburban, and rural environments.
The measurements were performed using an embedded \ac{V2X} platform with a direct modem interface, \emph{which allows to access the current transmission power} in order to obtain the ground truth label for the prediction. The device was placed in the rear trunk of a car and periodically uploaded files of \SI{1}{\mega\byte} to \SI{5}{\mega\byte} to a \ac{HTTP} server.
The integrated \LTE modem provides 31 network indicators including the current transmission power.
From the three considered machine learning models, \ac{RR}, \ac{DL}, and \ac{RF}, the latter achieved the lowest \ac{RMSE} of \SI{5}{\dB} to \SI{6}{\dB}, depending on the feature set.
In addition, the absolute sum of errors shrinks as the number of predictions grows and falls below \SI{1}{\dB} after \num{28} predictions.
Hence, this approach is well-suited for long-term applications and post-processing analysis of large data sets.

In this paper, the \ac{RF}-based approach for \emph{practical} applications is applied to predict the $\tilde{P}_{\text{TX}}$ for transmissions based on the indicators \RSRP, \RSRQ, \SINR, upload size, and the vehicle's velocity.
Finally, the dwell times of \ac{CoPoMo}'s four-state model are computed by mapping the $\tilde{P}_{\text{TX}}$ predictions into the corresponding power states, which in turn enables an estimation of the average power consumption of the \UE.

In conclusion, the presented process allows to analyze the energy-efficiency of the considered transmission schemes without the requirement for dedicated measurement equipment and explicitly without knowledge of the applied transmission power. Although this approach is utilized for offline result analysis in the next section, it can also be applied for online prediction directly on the embedded device.

\section{Empirical Results of Vehicular Sensor Data Transfer} \label{sec:results}

In this section, the results of the empirical performance evaluation are presented and discussed. At first, the impact of the different considered context information on the resulting uplink data rate is evaluated. Afterwards, results for \emph{data rate}, \emph{age of information} and \emph{uplink power consumption} are presented for the context-aware transmission scheme using \ac{CAT} with different single downlink indicators as transmission metrics. Then, the accuracy of the mobility prediction schemes and their impact on the predictability of the future context information is discussed and the results for the context-predictive scheme \ac{pCAT} are presented. Finally, detailed measurements for the machine learning-enabled transmission methods \ac{ML-CAT} and \ac{ML-pCAT} are provided.

\subsection{Correlation of Downlink Indicators and Data Rate}

\begin{figure*}[] 
	\centering
	\includegraphics[width=\triple]{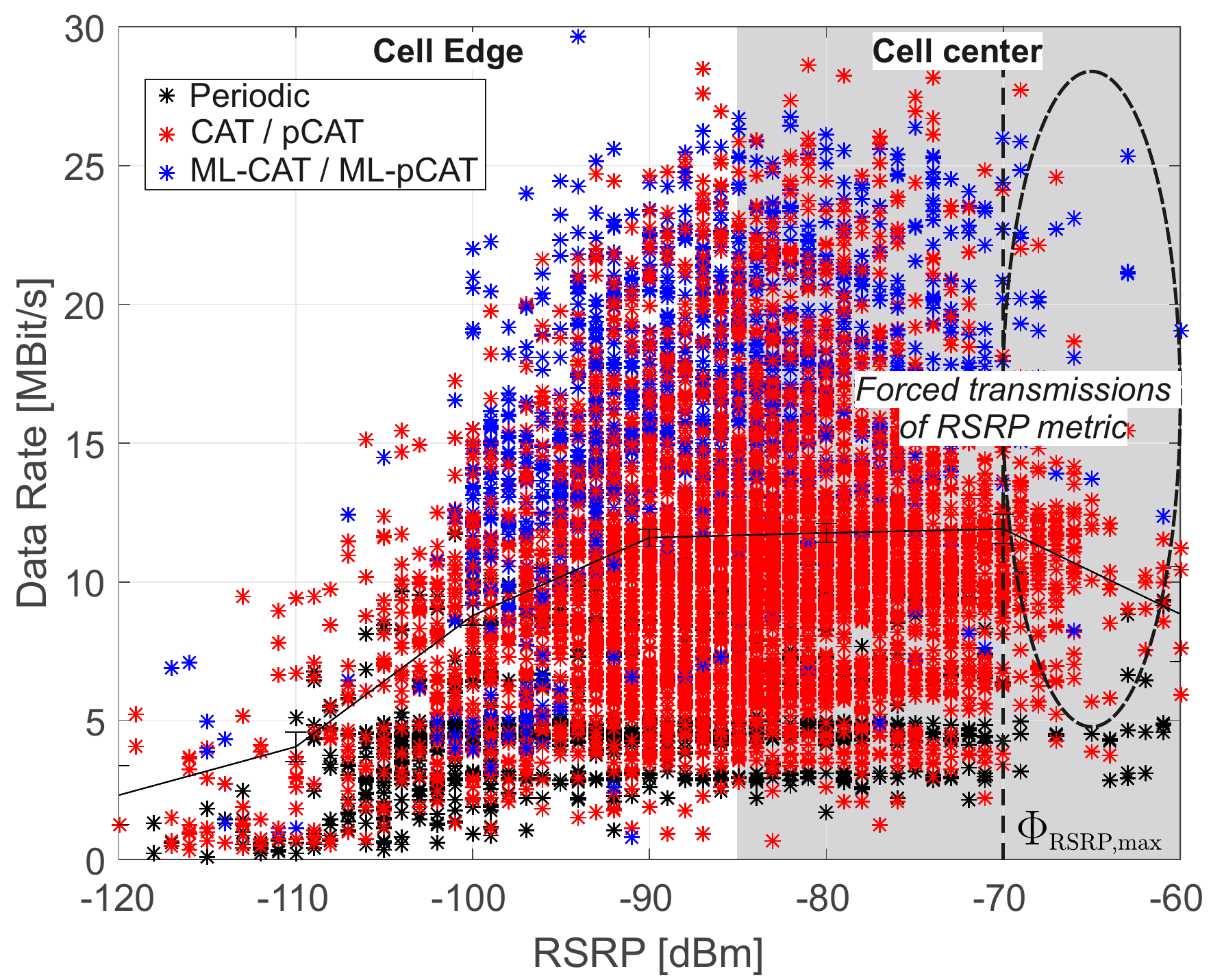}\quad
	\includegraphics[width=\triple]{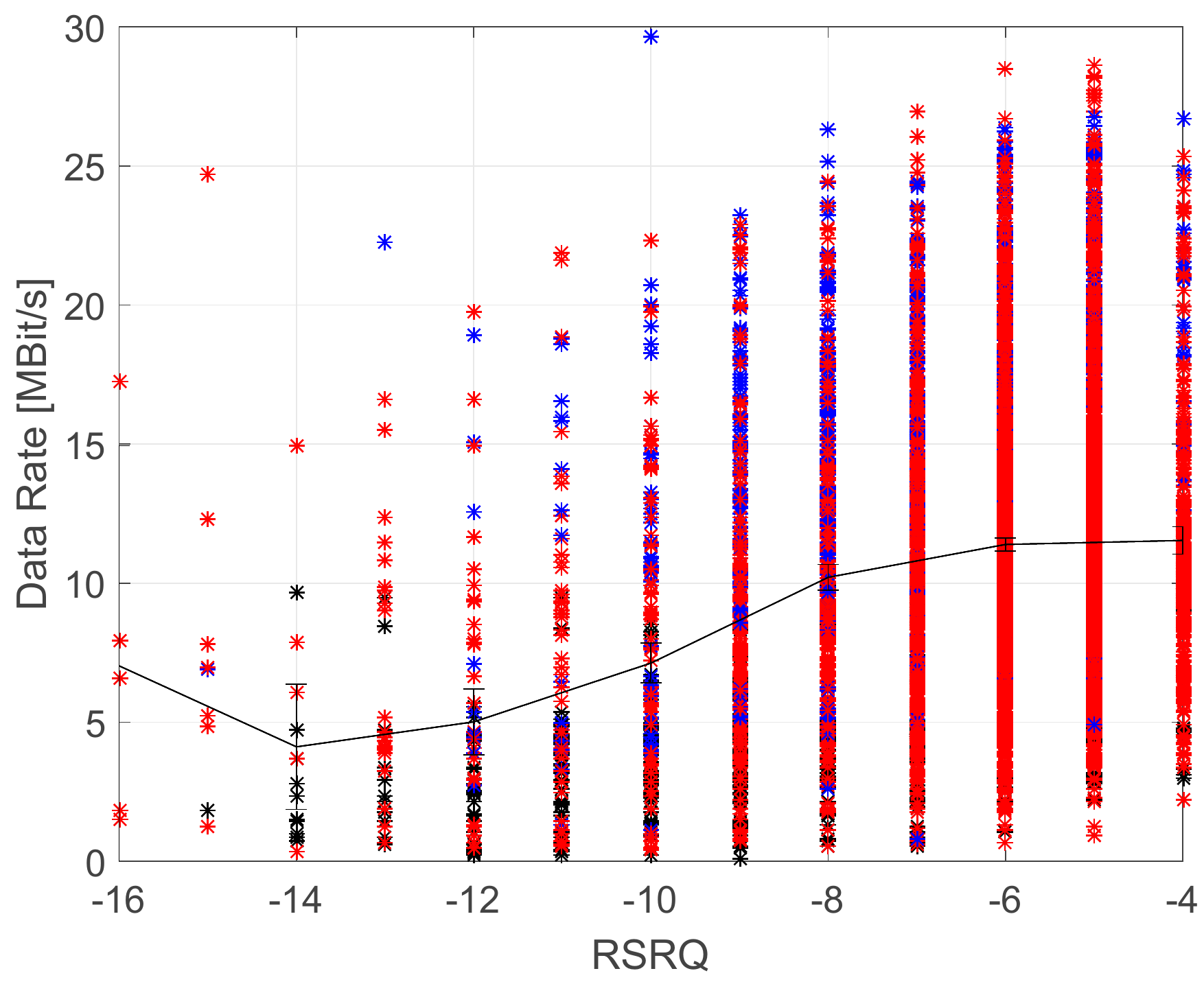}\quad
	\includegraphics[width=\triple]{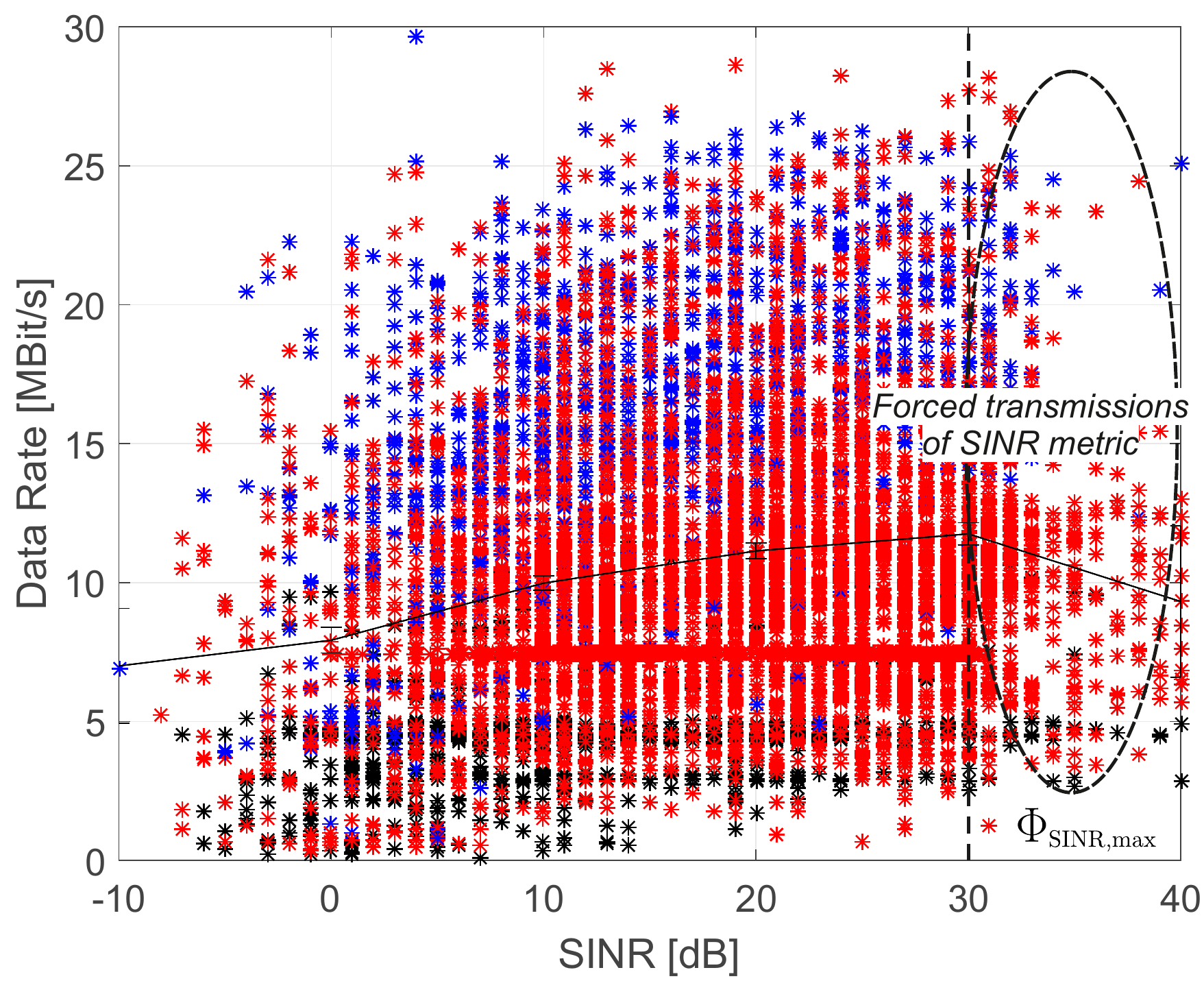}
	\medskip
	\includegraphics[width=\triple]{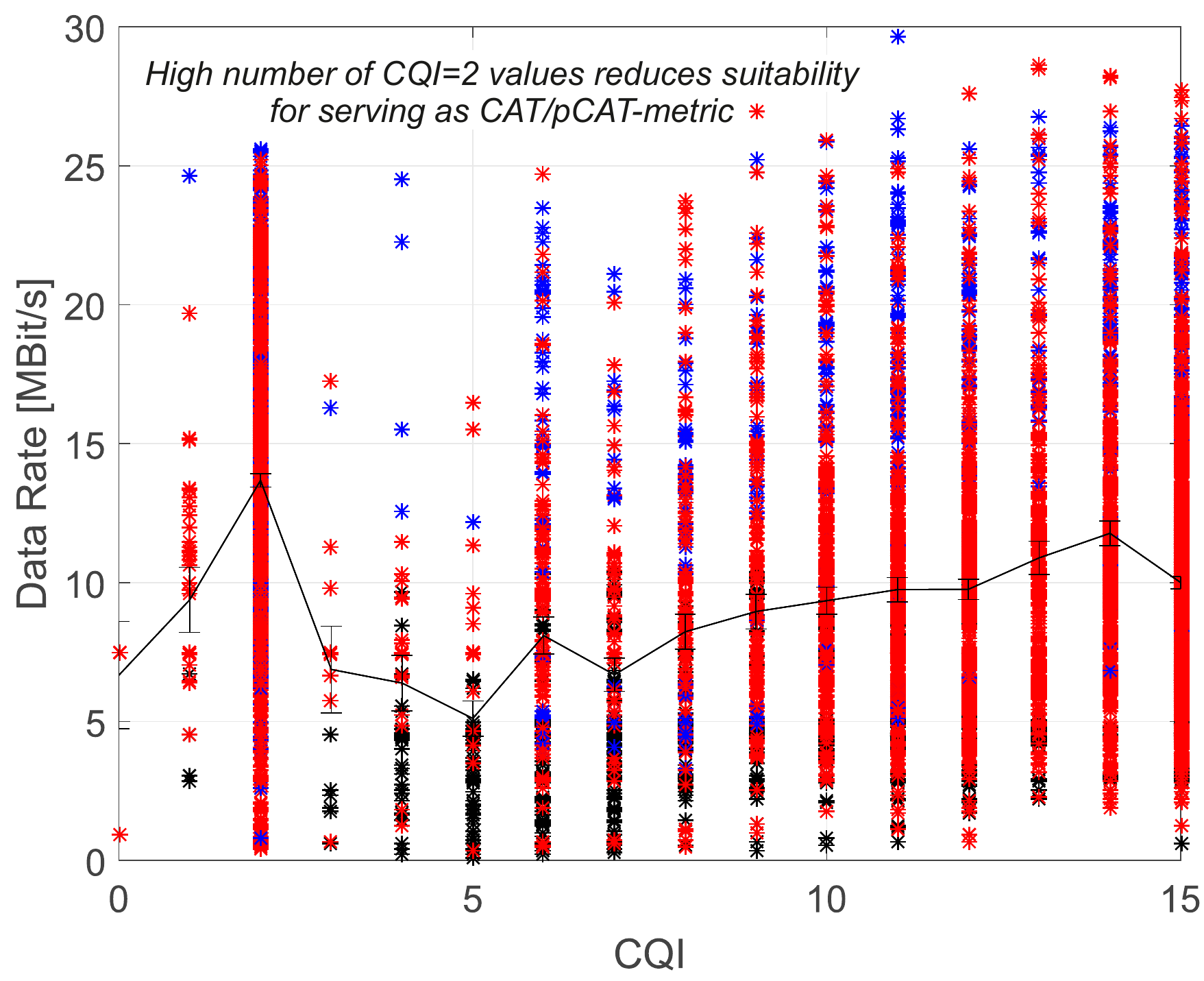}\quad
	\includegraphics[width=\triple]{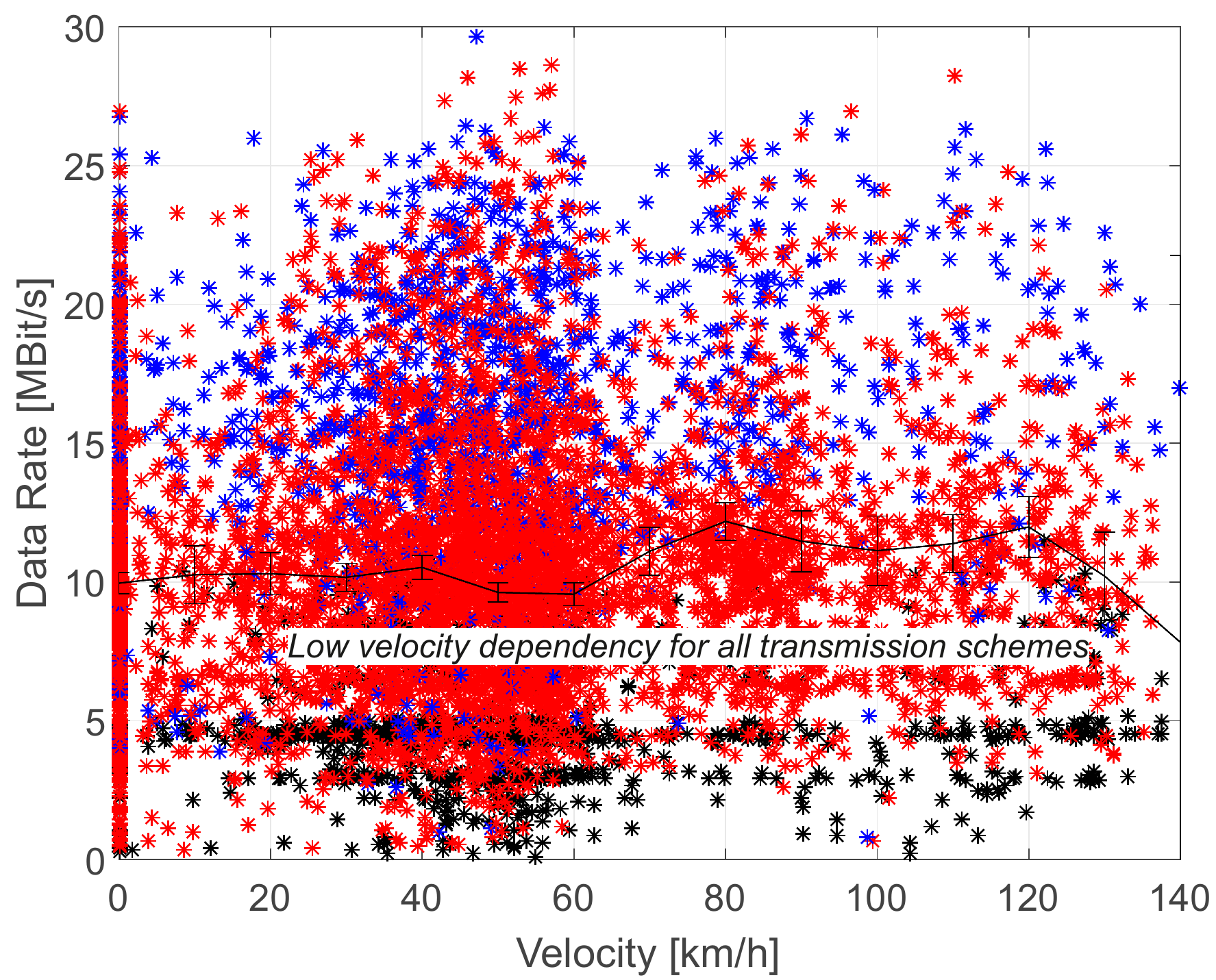}\quad
	\includegraphics[width=\triple]{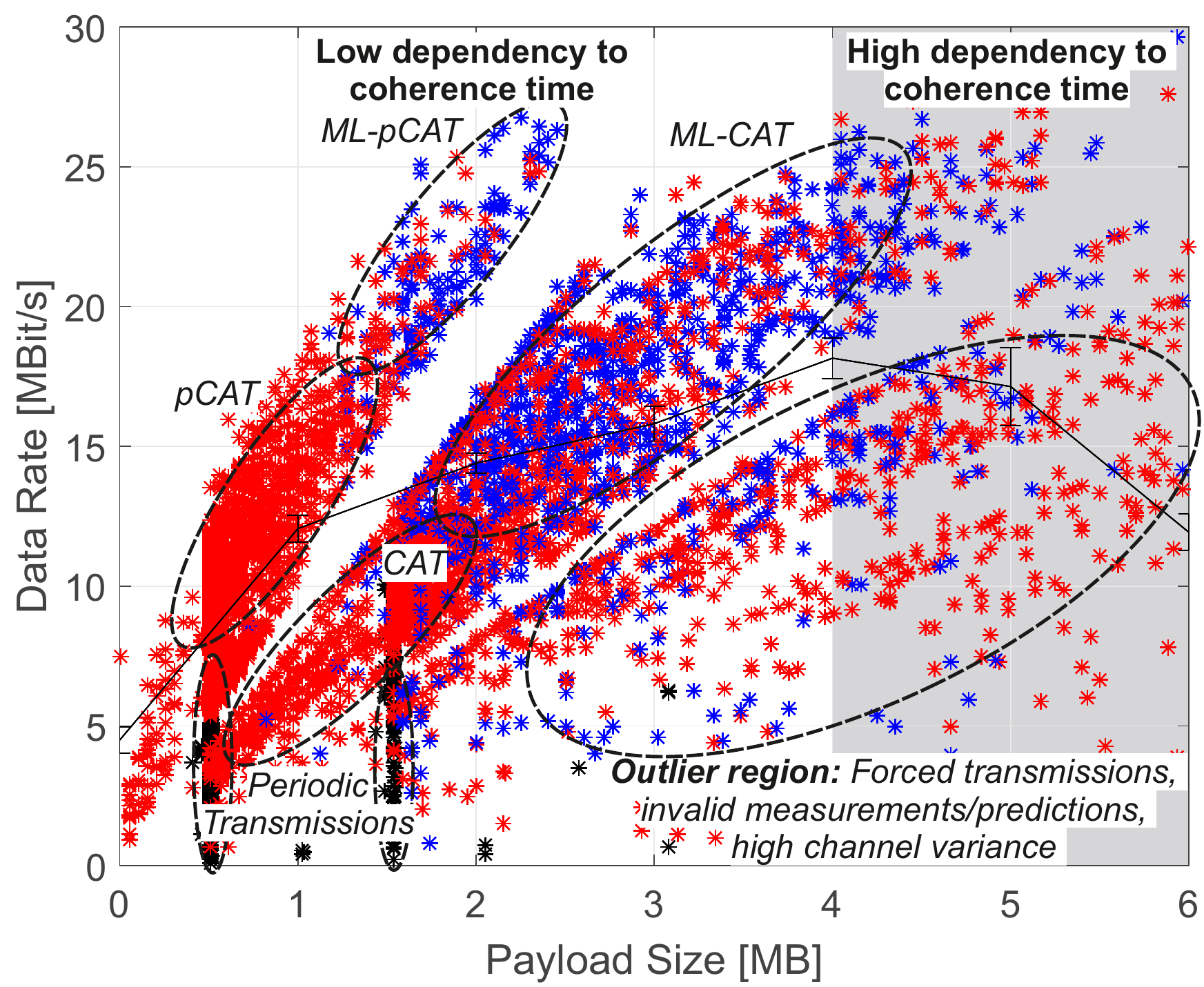}
	
	\vspace{-0.5cm}
	\caption{Correlation of different indicators with the resulting end-to-end data rate. The black curve shows the 0.95 confidence interval of the overall mean value. The plots consist of data from periodic (\emph{black}), \ac{CAT} and  \ac{pCAT} (red) and \ac{ML-CAT} and \ac{ML-pCAT} (blue) transmissions. Forced transmissions are caused for all \ac{CAT}-based approaches if the measured metric value $\Phi(t)$ exceeds the maximum defined metric value $\Phi_{\text{max}}$.}
	\label{fig:correlation}
	\vspace{-0.2cm}
\end{figure*}

The correlation of \ac{RSRP}, \ac{RSRQ}, \ac{SINR}, \ac{CQI}, velocity and payload size with the resulting data rate is shown in \Fig{\ref{fig:correlation}}. Since these indicators also are the features of the data rate prediction, the analysis gives an impression about the importance of the different features for the overall prediction behavior. The plots contain the individual transmissions of the whole data set, consisting of periodic, \ac{CAT} and \ac{pCAT} data transfer.
%
%
\\
%
%
It should be denoted that the resulting value range of the data rate exceeds the limits given by the data rate prediction shown in \Fig{\ref{fig:m5t_eval}}. The reason for this behavior is that the \ac{pCAT}-based and the machine learning-enabled schemes, which mainly achieve these high values, were not part of the used training set (see Sec.~\ref{sec:data_rate_prediction}). Furthermore, the plots contain multiple \emph{forced transmissions}, which are triggered for all \ac{CAT}-based approaches, if $\Phi(t)\geq\Phi_\text{max}$.
As expected, the data points for the periodic transmission scheme are \emph{uniformly distributed} among the value ranges of the respective metrics as the transmissions are performed regardless of the channel quality.
%
%
The characteristics of the \ac{RSRP} can be divided into two distinct areas that are divided by a breakpoint at \SI{-85}{dBm}, that allows a division into \emph{cell edge}- and \emph{cell center}-behavior. In the \emph{cell edge} area, the \ac{RSRP} is a dominant factor for the achievable data rate, which is increased with higher \ac{RSRP} values. Within the \emph{cell center}, the dependency is decoupled since other effects (e.g. interference) have a more dominant influence on the behavior.
%
%
%
%
%
%
The behavior of the \ac{CQI} shows a peak for $\text{CQI=2}$. During the drive tests, those values occured frequently on both tracks and without any obvious correlation to the other indicators. For \ac{LTE}, the actual calculation of this indicator is not standardized and depends on the modem manufacturer. It can be concluded that the reported \ac{CQI} is limited for being used as a \ac{CAT}-metric, which is also confirmed by the evaluations in the following sections.
\\
%
%
Although the drive tests were performed within a velocity range of \SIrange{0}{140}{\kilo\meter\per\hour}, the observed dependency of the data rate to the velocity is very low.
%
%
\\
For the payload size, multiple lobes can be identified that separate periodic transmissions, \ac{CAT}, \ac{pCAT}, \ac{ML-CAT} and \ac{ML-pCAT}. Another region mainly consists of outliers that are related to forced transmissions either caused by $\Phi_{\text{max}}$ or $t_{\text{max}}$ as well as inaccurate measurements and high channel variances. 
%
%
It can be observed that the machine learning approaches are systematically able to achieve higher data rates for the same payload size. Especially the introduced look-ahead of \ac{ML-pCAT} approaches is able to proactively avoid transmissions during low channel quality periods. This fact is underlined by the correlation analysis in \cite{Sliwa/etal/2018b}, that is based on an earlier version of the data set and does not contain the \ac{pCAT}- and \ac{ML-pCAT}-specific lobes.
%
%
The value range for the payload size is limited to \SI{6}{\mega\byte} as $t_{\text{max}}$ is defined as \SI{120}{\second} and the sensor application generates \SI{50}{\kilo\byte} of data per second.
%
%
The overall behavior can be characterized by two areas that have different grades of dependencies to the channel coherence time. Up to \SI{4}{\mega\byte}, the data rate highly benefits from increased payload sizes as the slow start of \ac{TCP} is less dominant for the overall transmission duration and a better payload-overhead-ratio is achieved. After the breakpoint, the probability for low data rate transmissions is highly increased as the channel is more likely changing its characteristics during active transmissions due to the longer transmission duration.
\\
%
%
The correlation analysis shows that no single indicator is able to provide a robust measurement for the channel quality in all considered situations.

\subsection{Single-metric Context-aware Transmission}
\vspace{-0.1cm}

\begin{figure*}[] 
	\centering
	
	\includegraphics[width=\triple]{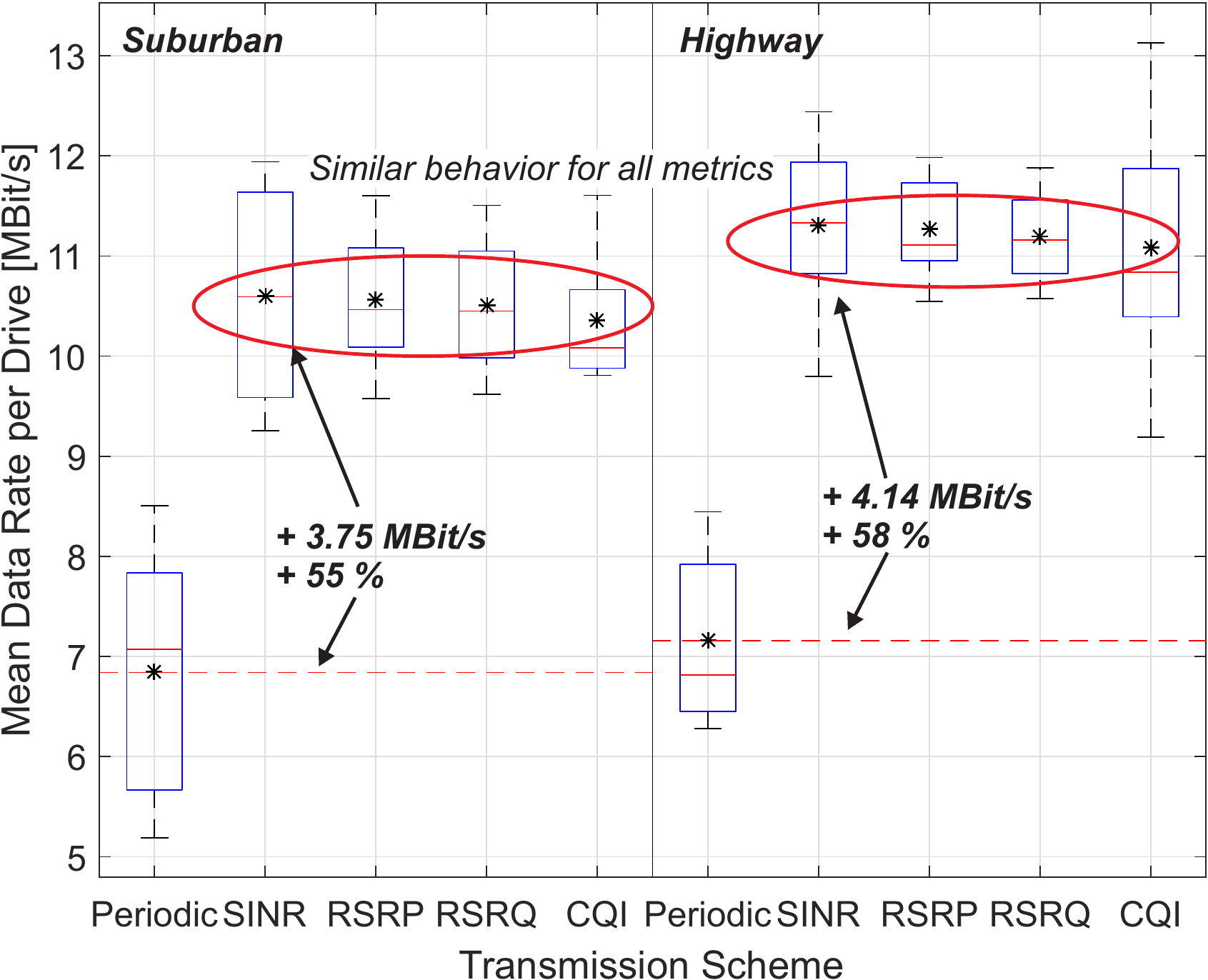}\quad
	\includegraphics[width=\triple]{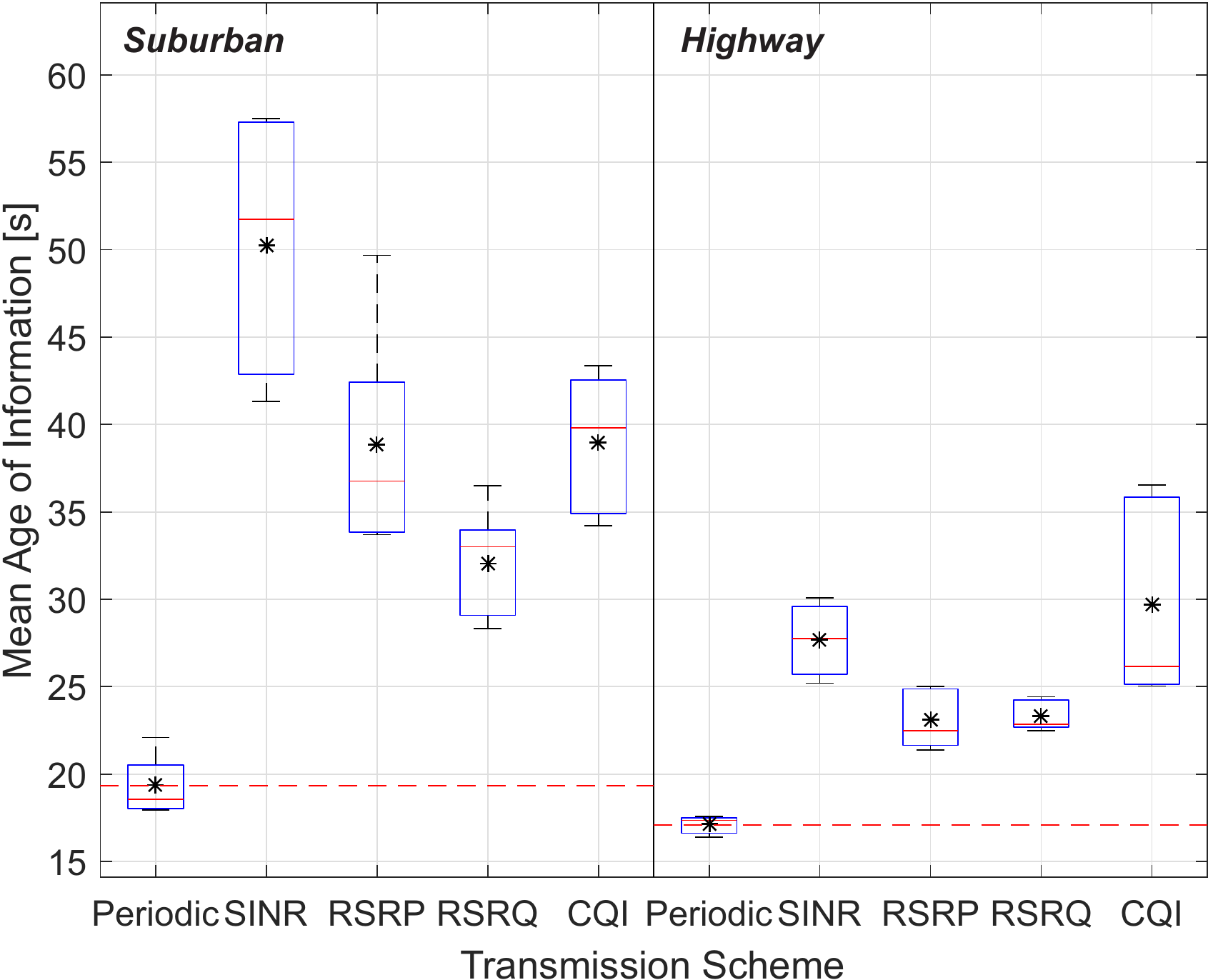}\quad
	\includegraphics[width=\triple]{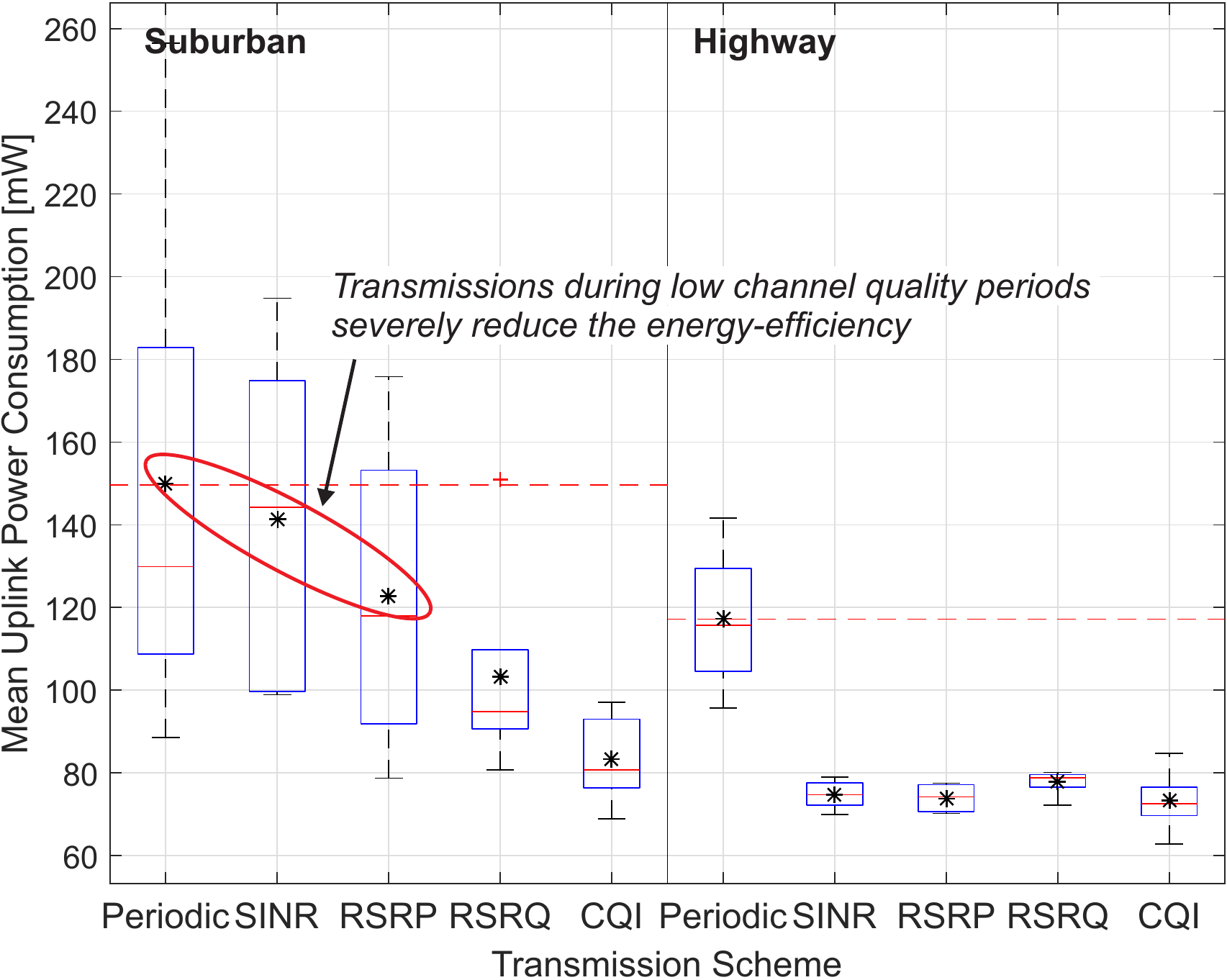}
	
	\vspace{-0.2cm}
	\caption{Performance comparison for the context-aware transmission scheme \ac{CAT} with different single-indicator metrics $\Phi$. The results for periodic transmission with fixed interval of \SI{30}{\second} are shown as reference.}
	\label{fig:box_cat_single}
	\vspace{-0.5cm}
\end{figure*}

The results for the context-aware transmission are shown in \Fig{\ref{fig:box_cat_single}}. Multiple variants of \ac{CAT} are configured with each of the passive downlink indicators as a metric according to \Tab{\ref{tab:metrics}.} The results for periodic transmission with a fixed interval of \SI{30}{\second} are shown as reference.
%
%
%
%
Using the context-aware approach, an average data rate gain of \SI{55}{\percent} is achieved for \ac{CAT}-based metrics with $\Phi_{\text{SINR}}$ and $\Phi_{\text{CQI}}$ having the highest variance during the different evaluation tests.
%
%
Although the results for the data rate are similar for all \ac{CAT}-based metrics, the \ac{AoI} behavior shows significant differences, which are related to the dynamics of the corresponding network quality indicator during the drive tests. Here, high \ac{AoI} values are an indicator for longer periods of low channel quality that prevent the vehicle from transmitting its data. In the suburban scenario, the \ac{SINR} values are rarely close to $\Phi_\text{max}$ and even multiple transmissions are forced by the maximum buffering delay $t_{\text{max}}$, resulting in a very high \ac{AoI}. All schemes highly exceed the baseline defined by the periodic transmission approach. Yet, the up-to-dateness of sensor measurements is still sufficient for the considered crowdsensing scenario (see \Sec{\ref{sec:introduction}}). On the highway track, the average \ac{AoI} is reduced, as the measurement channel behavior is frequently changing due to the high velocity of the vehicle, resulting in a higher transmission frequency.
%
%
These aspects are further confirmed by analyzing the uplink power consumption. It can be seen that transmissions during low quality periods --- here most clearly illustrated by the $t_{\text{max}}$-related forced transmissions for the \ac{SINR}-metric --- severely increase the average power consumption and reduce the energy-efficiency.

\subsection{Impact of Mobility Prediction on the Network Quality Indicators} \label{sec:mobility_prediction_accuracy}
\vspace{-0.1cm}

Before the measurement results of the context-predictive transmission schemes are presented, the accuracy of the mobility prediction mechanisms and the impact of prediction errors on the channel context estimation is discussed.
%
%
\begin{figure*}[]  
	\centering
	
	\includegraphics[width=\triple]{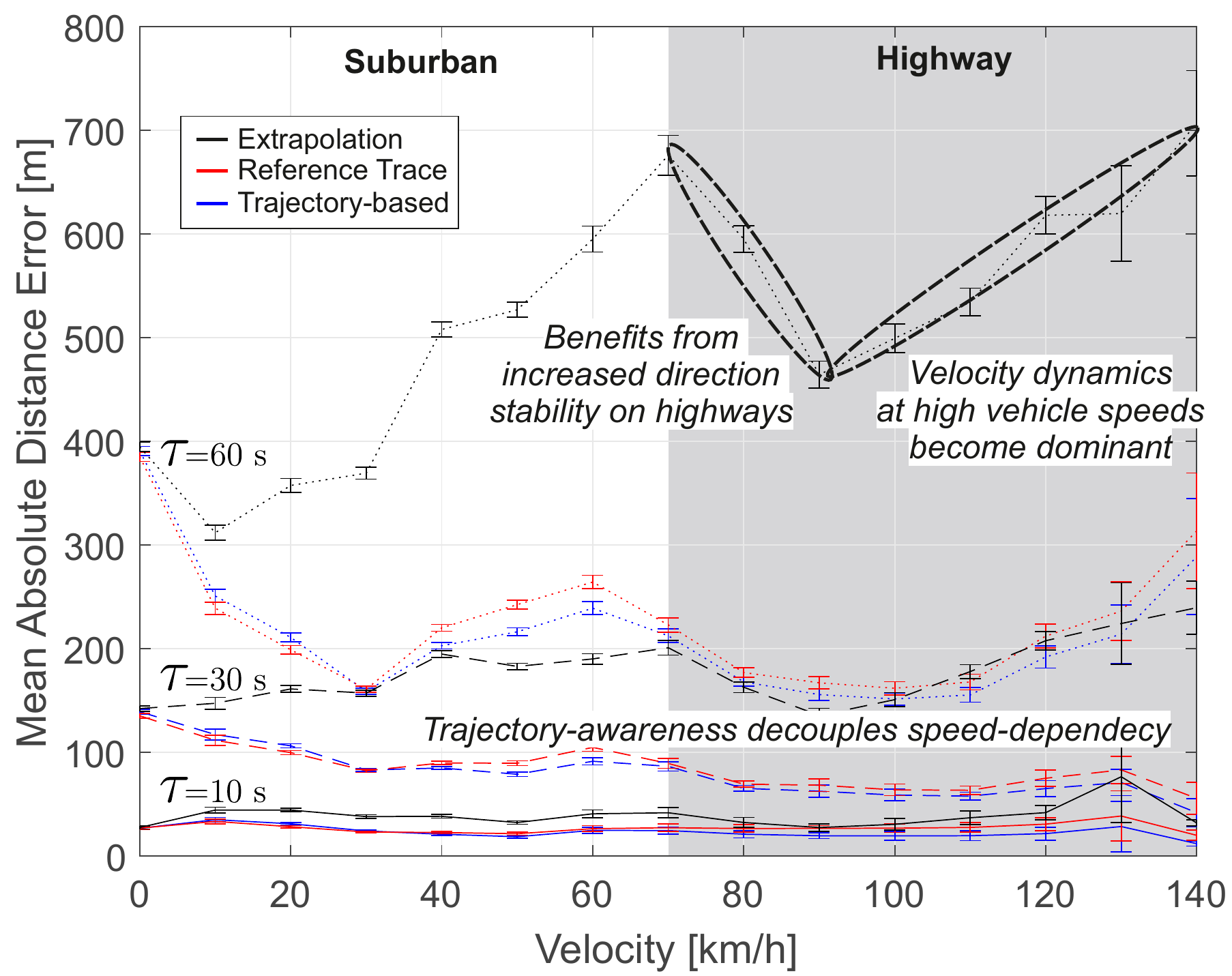}\quad
	\includegraphics[width=\triple]{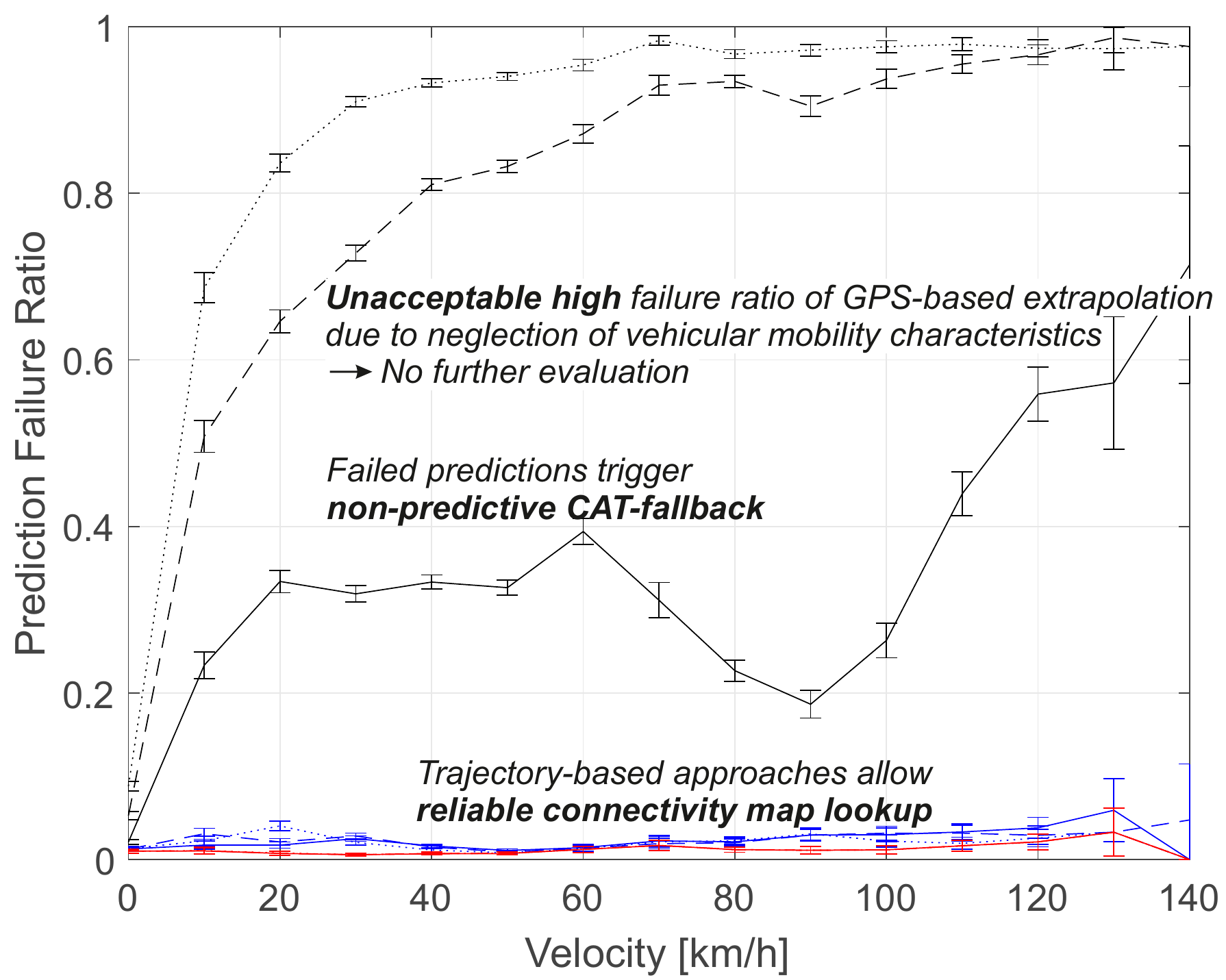}\quad
	\includegraphics[width=\triple]{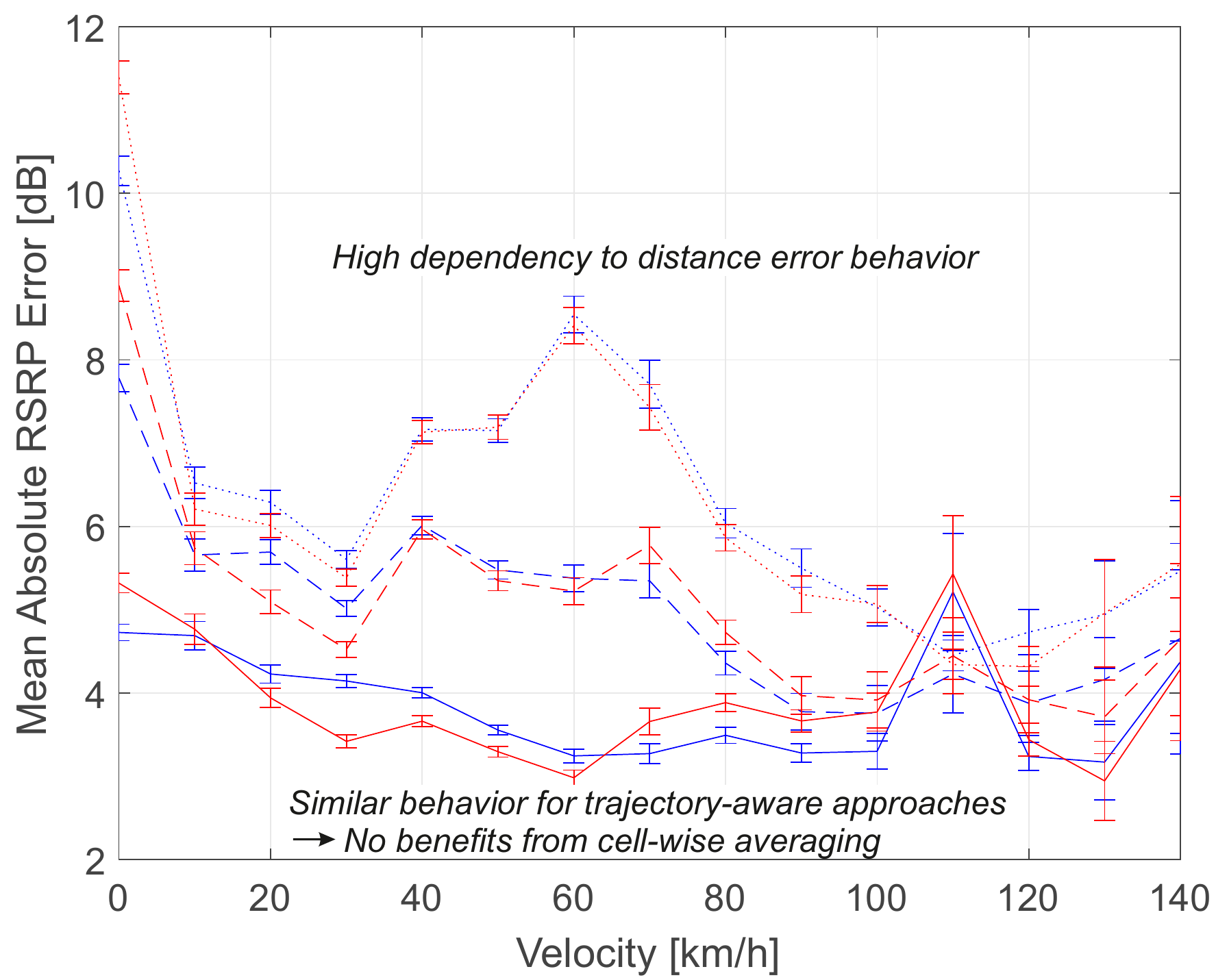}
	
	\medskip
	
	\includegraphics[width=\triple]{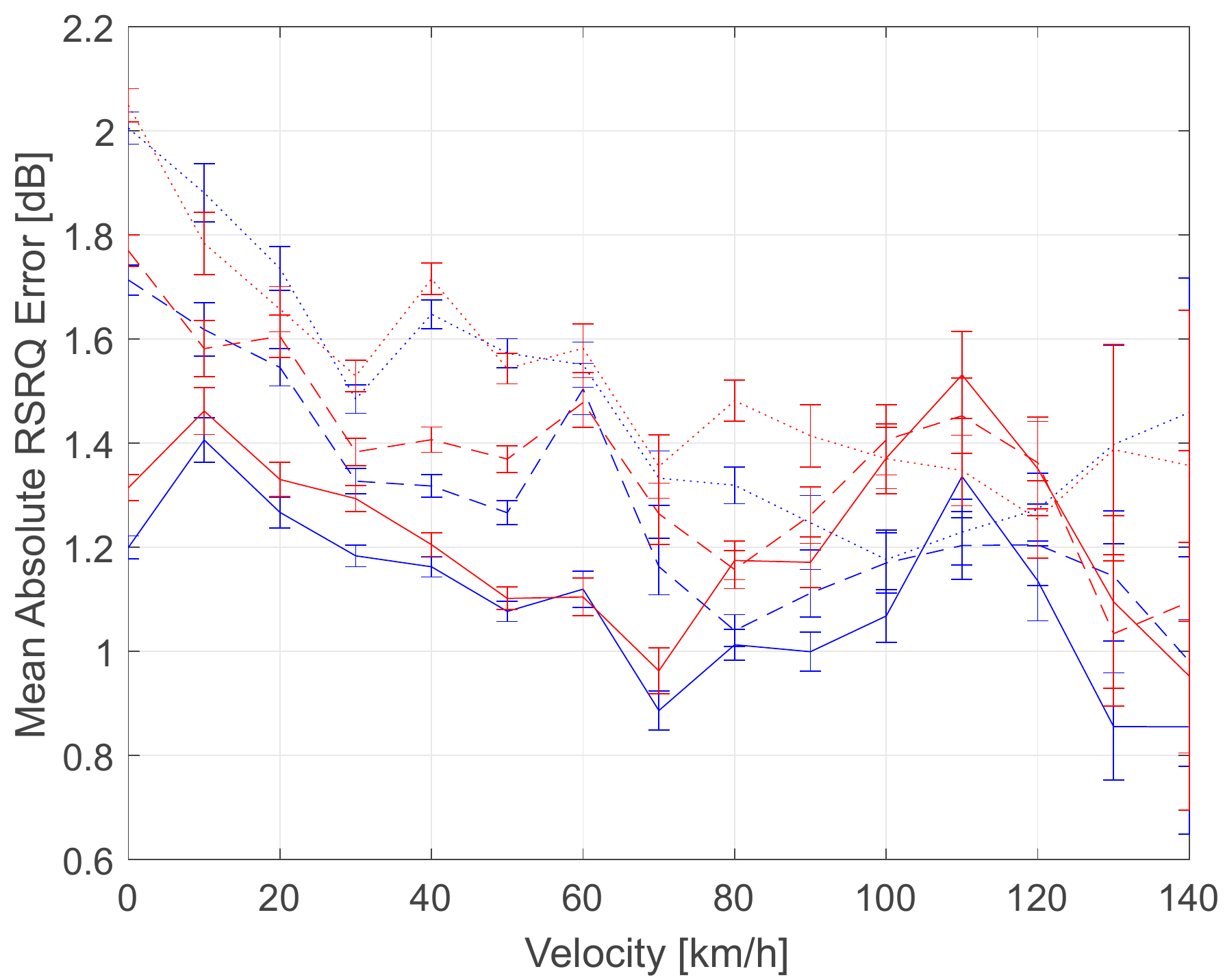}\quad
	\includegraphics[width=\triple]{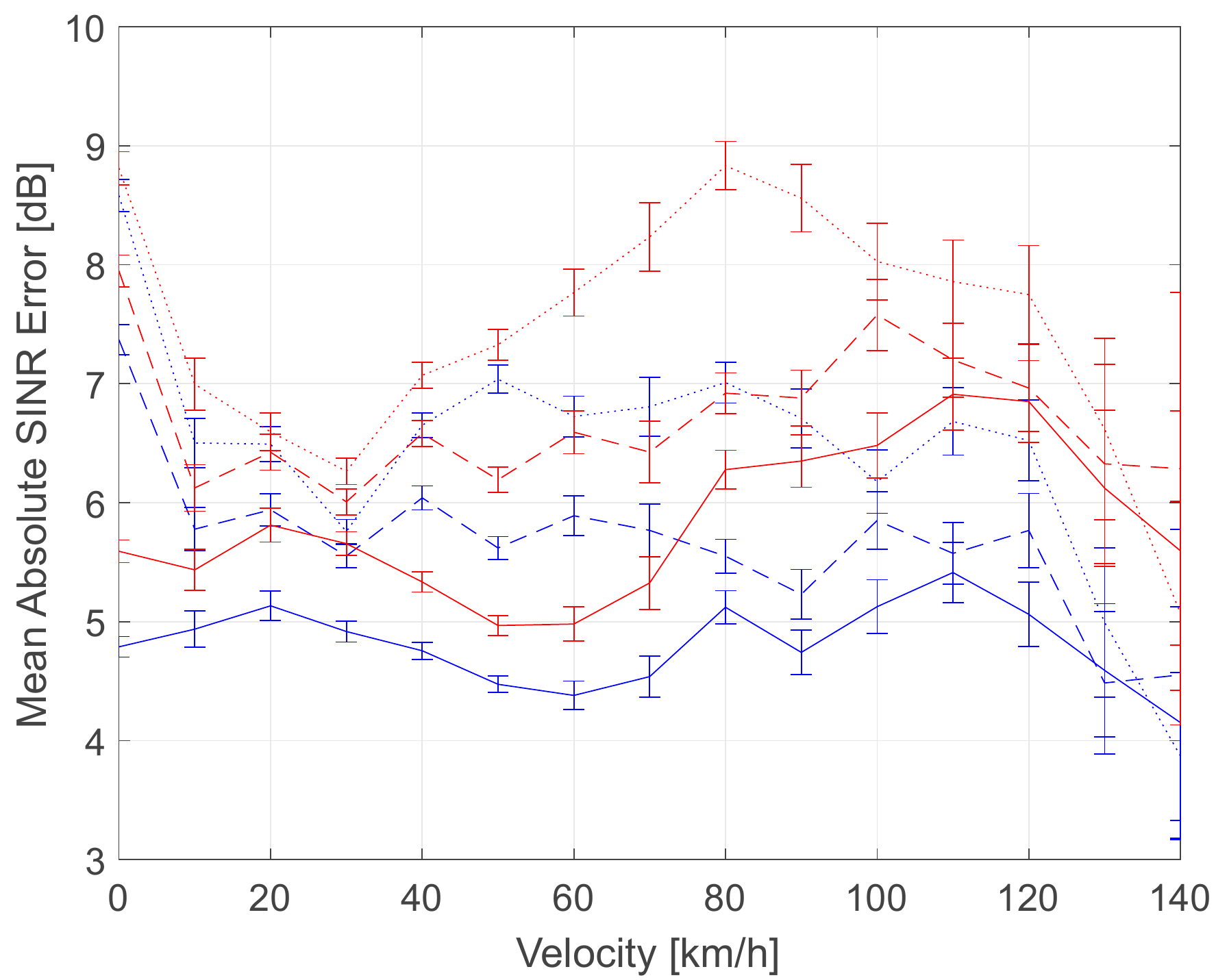}\quad
	\includegraphics[width=\triple]{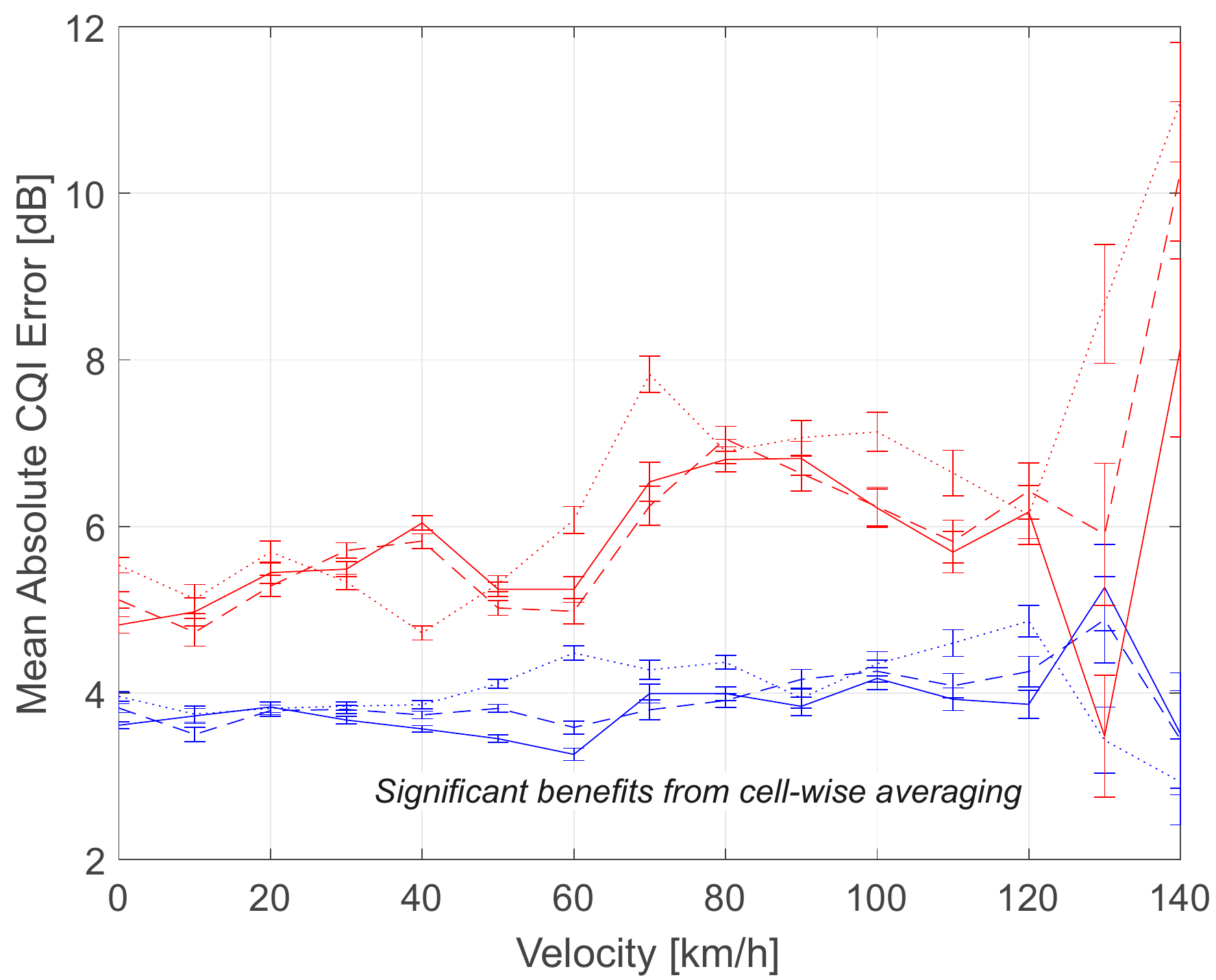}

	\vspace{-0.2cm}
	\caption{Prediction accuracy for the vehicle's mobility and the passive downlink indicators of the \ac{LTE} signal for the considered prediction methods with different predictions horizons in relationship to the vehicle's velocity. All curves show the 0.95 confidence interval of the mean value.}
	
	\label{fig:mobility_prediction_accuracy}
	\vspace{-0.6cm}
\end{figure*}
An evaluation of the velocity-dependent mobility prediction accuracy as well as the implications for error-effected forecasts on the network quality assessment is provided in \Fig{\ref{fig:mobility_prediction_accuracy}}. 
%
%
The accuracy of the \ac{GPS} extrapolation approach is significantly influenced by the probability of direction changes during the prediction horizon $\tau$. With regard to the speed-dependency of the prediction error, three characteristical regions can be identified. Up to \SI{70}{\kilo\meter\per\hour} (urban/suburban roads), the error dimension is proportional to the velocity. For higher velocities, the vehicle is more likely moving on a highway track with a low probability for direction changes. However, above \SI{90}{\kilo\meter\per\hour} the higher prediction distance becomes the dominant error source again. 
For the trajectory-aware approaches, the resulting distance error is much lower. Moreover, due to the consideration of the vehicle's turn behavior, the dependency between prediction accuracy and velocity is decoupled, allowing robust predictions even for higher values of $\tau$.
%
%
The future position $\tilde{\vec{P}}(t+\tau)$ is used to look up the future channel context $\vec{C}(t+\tau)$ from the connectivity map. Therefore, inaccurate forecasts may lead to situations where the connectivity map does not contain data for the (falsely) predicted cell, which is statistically captured by the \ac{PFR} metric. For the \ac{pCAT}-based schemes, prediction failures trigger the \emph{\ac{CAT}-fallback}, where the respective transmission scheme behaves equally to a pure probabilistic \ac{CAT} scheme with the same metric properties.
%
%
As a consequence, only the trajectory-based prediction methods are further considered in the following as the \emph{\ac{PFR} is unacceptably low for the extrapolation approach}.
%
%
For assessing the added information by predicting the passive downlink indicators, the resulting error has to be set into relation to the value range $\Phi_{\textbf{max}}-\Phi_{\textbf{min}}$ of the corresponding metric $\Phi$ and is severely influenced by the accuracy of the position prediction.
While for \ac{RSRP} and \ac{RSRQ} only slight differences between the two prediction approaches can be observed, \ac{SINR} and \ac{CQI} achieve an \emph{aggregation gain} by the cell-wise averaging within the connectivity maps. As both indicators are affected by short-term fading, the single reference trace is not able to provide an accurate estimation.

\subsection{Single-metric Context-predictive Transmission} 

As a consequence of the prediction accuracy analysis, the context-predictive \ac{pCAT} scheme uses the trajectory-based mobility prediction with connectivity maps.
%
%
\begin{figure*}[] 
	\centering
	
	\includegraphics[width=\triple]{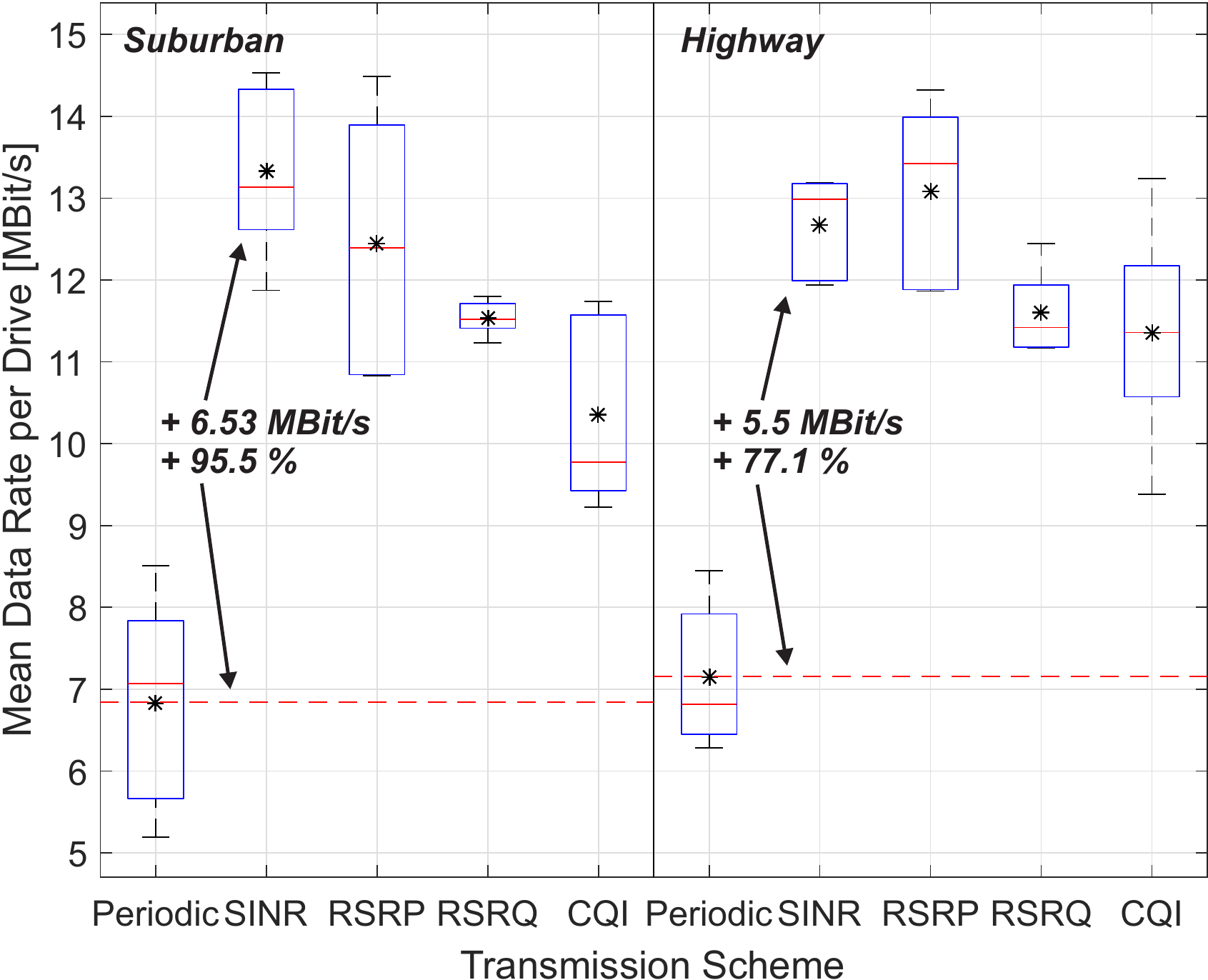}\quad
	\includegraphics[width=\triple]{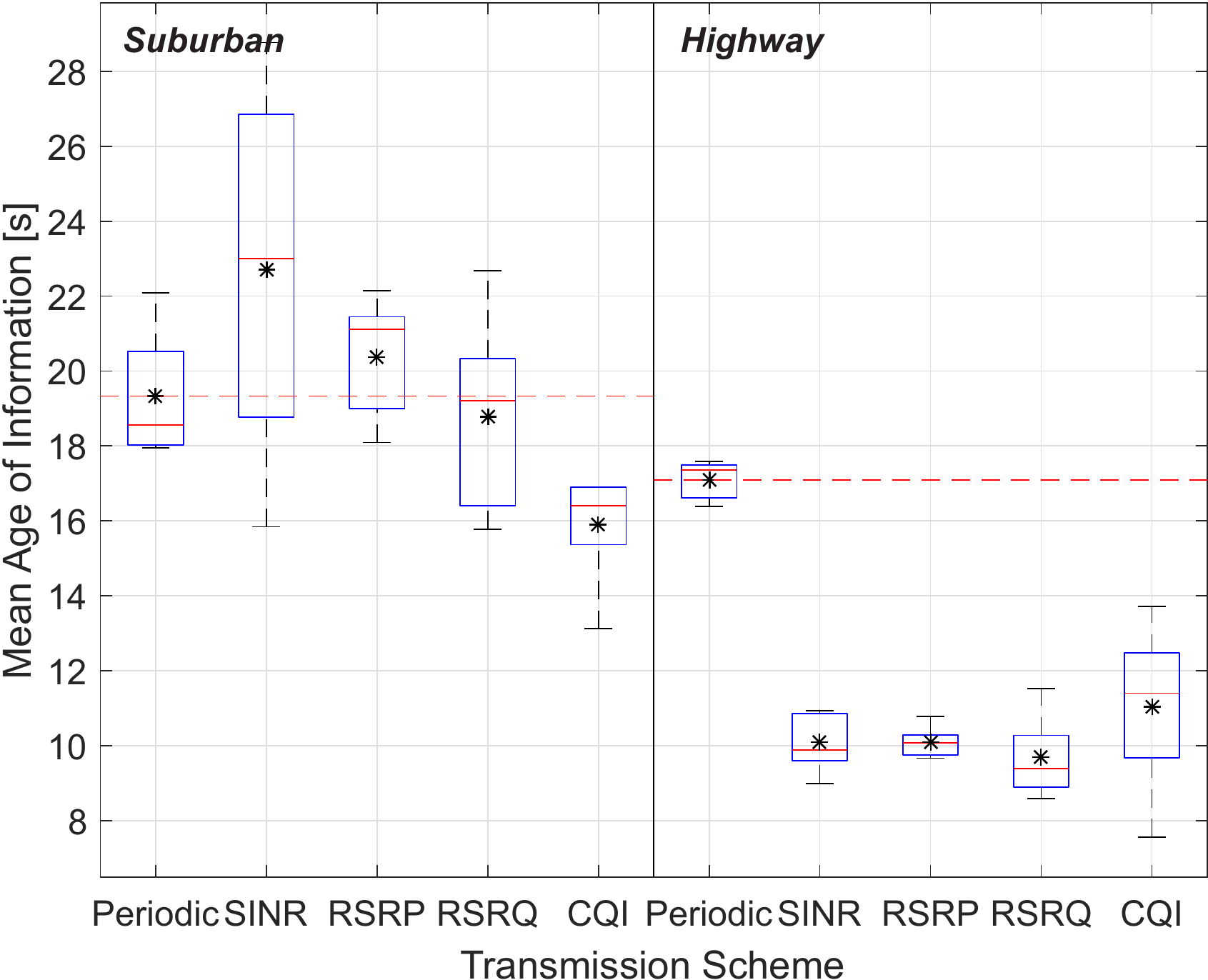}\quad
	\includegraphics[width=\triple]{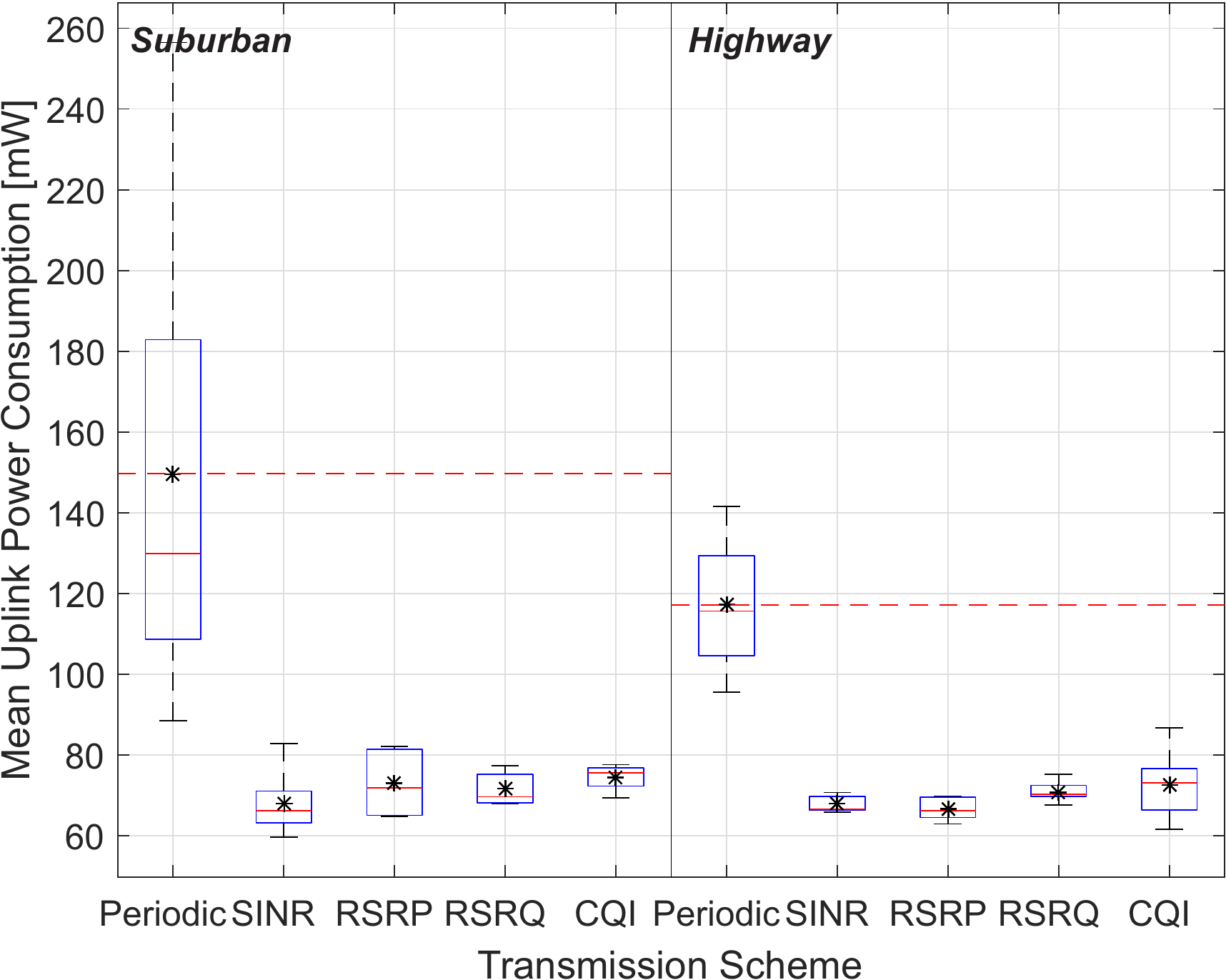}
	
	\vspace{-0.2cm}
	\caption{Performance comparison for the context-predictive transmission scheme \ac{pCAT} with $\tau=30~s$ and different single-indicator metrics $\Phi$.}
	\label{fig:box_pcat_single}
	\vspace{-0.2cm}
\end{figure*}
%
%
\Fig{\ref{fig:box_pcat_single}} shows the results for the considered \acp{KPI} with prediction horizon $\tau=\SI{30}{\second}$. By consideration of the future channel behavior, \ac{pCAT} adds another dimension that is sensitive to the dynamics of the channel quality and significantly changes the transmission behavior. With the vehicle proceeding on its route, the channel context is considered with a \emph{moving window} with the range $\left[t, t+\tau\right]$. As a consequence, the transmission scheme is much more influenced by the probability of the vehicle encountering the anticipated context within the remaining time interval to $t_{\text{max}}$ than on having a perfect prediction for a discrete point in time. 
%
%
Additionally, transmissions are performed more often, resulting in a reduced \ac{AoI}, which is even falling below the baseline of the periodic approach on the highway track.
\\
%
%
In contrast to \ac{CAT}, the resulting data rate is significantly different for the considered metrics as it is now also depending on the predictability of the metrics themselves. The highest gains are achieved with the \ac{SINR}- and the \ac{RSRP}-metrics (up to \SI{95}{\percent} on the suburban track and up to \SI{77}{\percent} on the highway track).
%
%
Due to the doubled dependency to the channel dynamics by consideration of $\vec{C}(t)$ and $\tilde{\vec{C}}(t+\tau)$ and the proactive detection of connectivity valleys, transmissions are less likely forced by $t_{\text{max}}$. As a consequence, the power consumption behavior of \ac{pCAT} significantly outperforms the \ac{CAT}-based approach.

\subsection{Machine Learning-enabled Transmission}

\begin{figure*}[] 
	\centering
	
	\includegraphics[width=\triple]{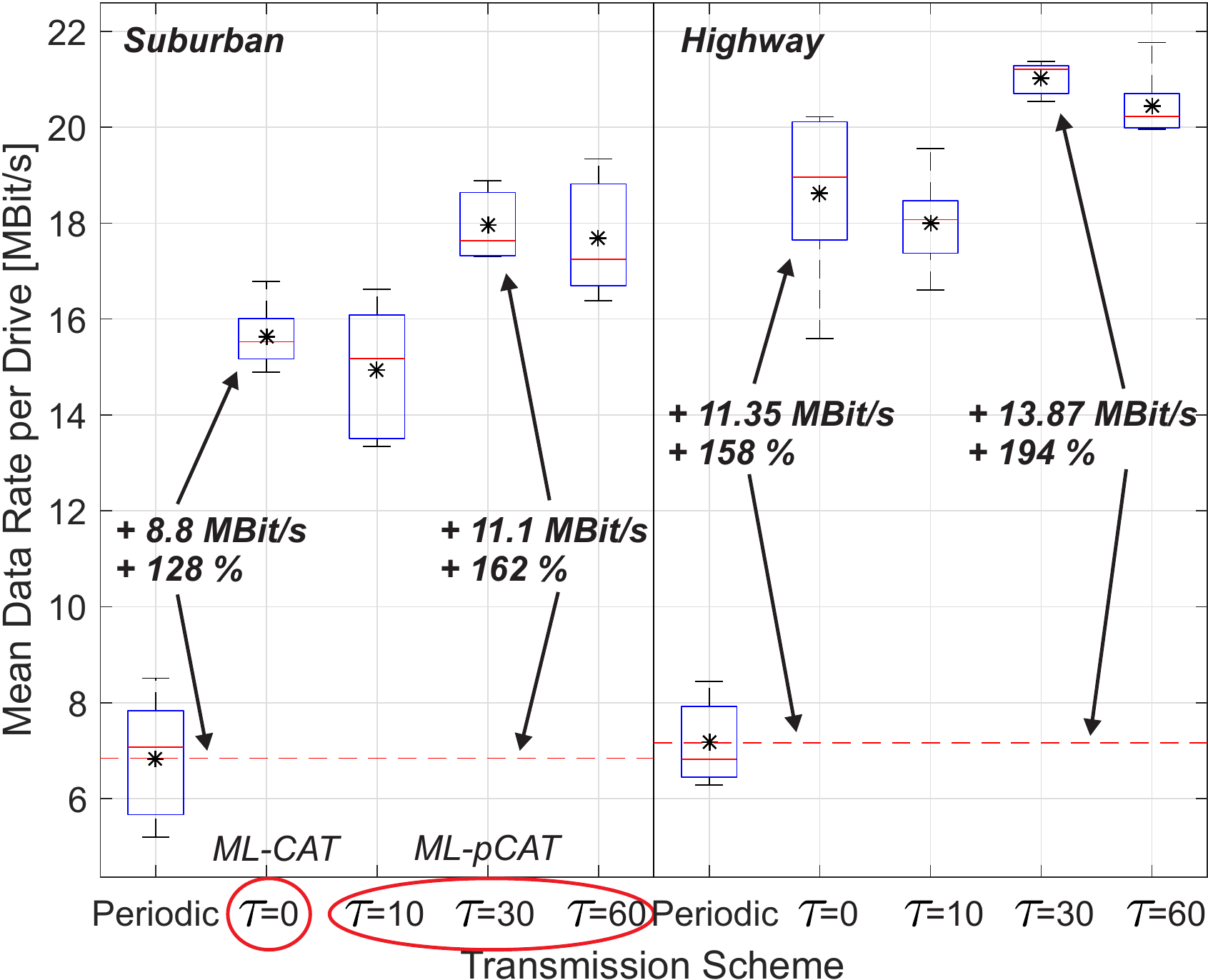}\quad
	\includegraphics[width=\triple]{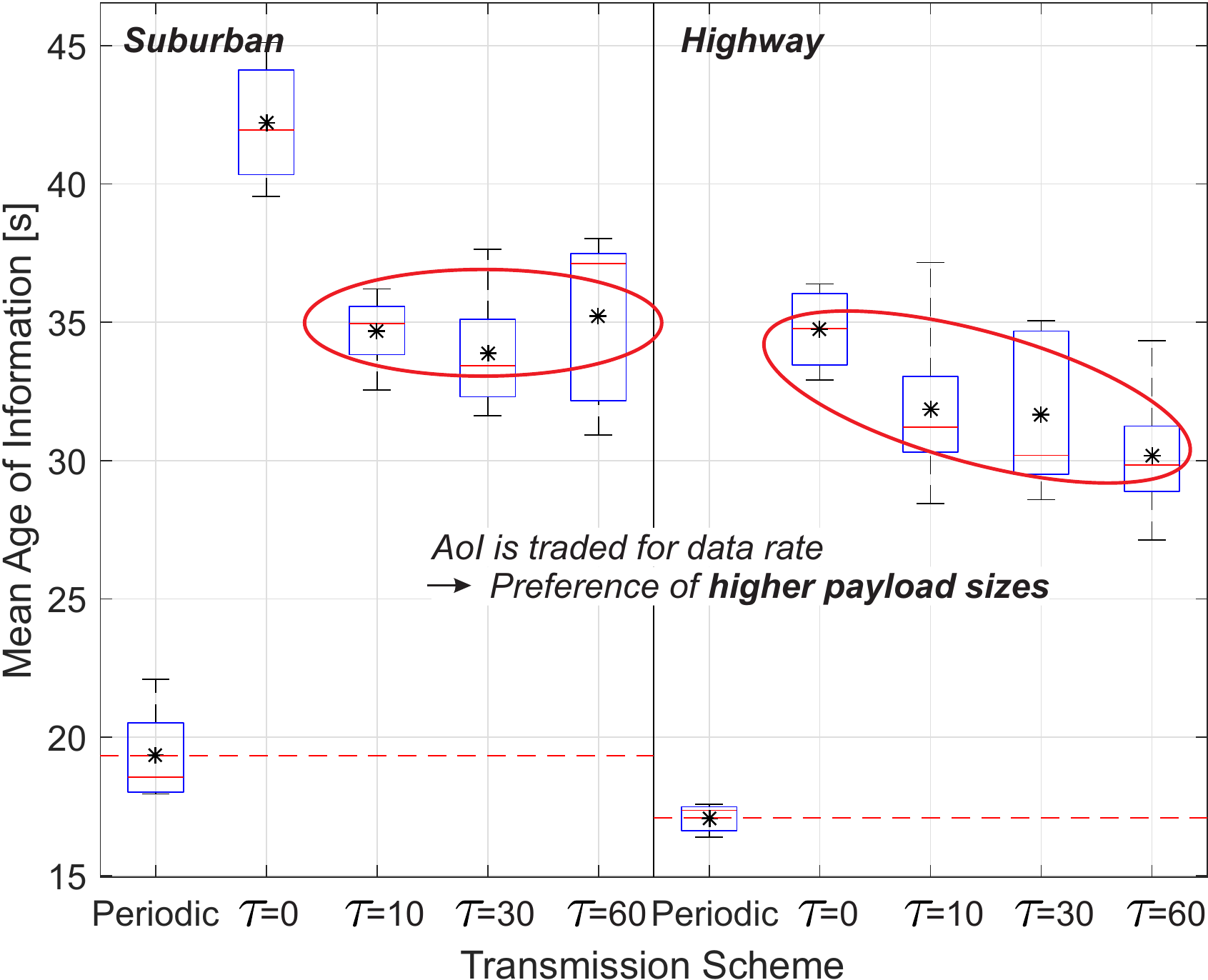}\quad
	\includegraphics[width=\triple]{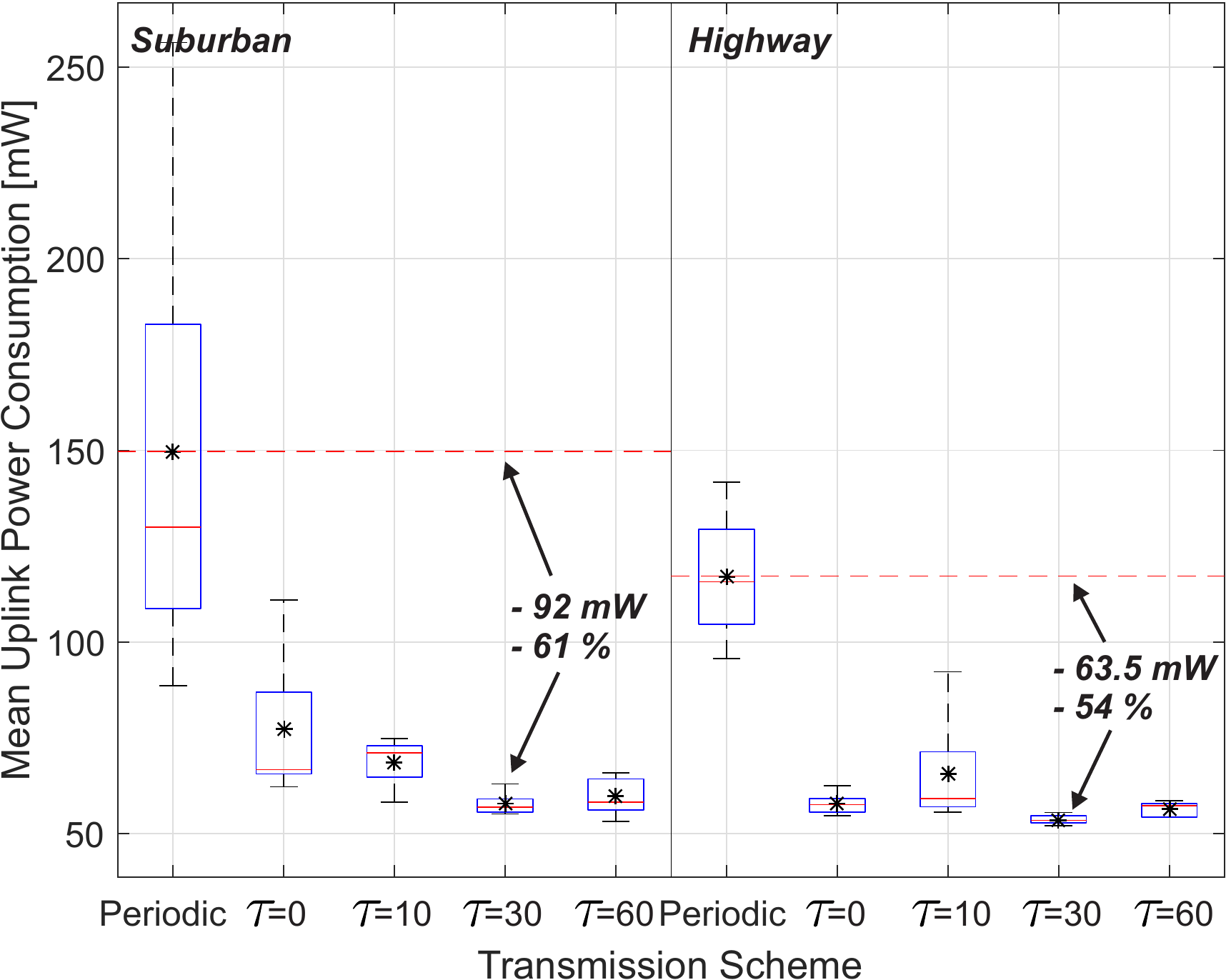}
	
	\vspace{-0.2cm}
	\caption{Performance comparison of the machine learning-enabled transmission schemes \ac{ML-CAT} ($\tau=\SI{0}{\second}$) and \ac{ML-pCAT} ($\tau=\left\lbrace 10, 30, 60 \right\rbrace$\si{\second}) using the $\Phi_{M5T}$ metric.}
	\label{fig:box_ml}
	\vspace{-0.5cm}
\end{figure*}

Finally, the results for \ac{ML-CAT} and \ac{ML-pCAT} are shown in \Fig{\ref{fig:box_ml}}.
%
%
The machine learning-enabled schemes exploit the correlation of the individual features with the resulting data rate as shown in \Fig{\ref{fig:correlation}} with the aim of data rate optimization.
%
%
Both variants are able to achieve massive boosts in the resulting data rate. The best overall performance is achieved for \ac{ML-pCAT} with $\tau=\SI{30}{\second}$ on the highway track, where it is able to achieve a data rate gain of \SI{194}{\percent} while simultaneously reducing the average uplink power consumption by \SI{54}{\percent}.
%
%
In comparison to the previously discussed results for single-metric \ac{pCAT}, the \ac{AoI} is increased as \ac{ML-CAT} and \ac{ML-pCAT} highly exploit the correlation between payload size and data rate. Therefore, these approaches actively introduce additional buffering delays to achieve higher packet sizes. 
%
%
\basicFig{}{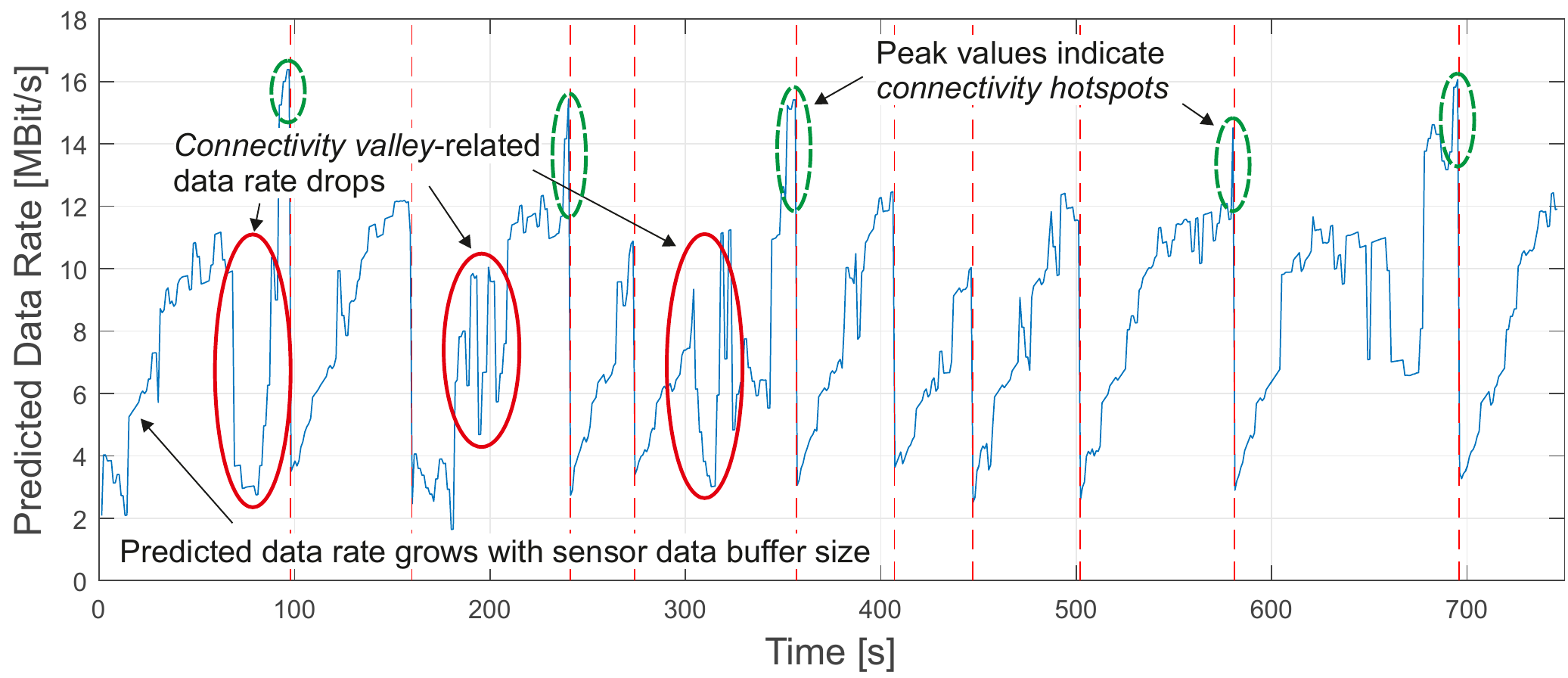}{Example temporal behavior of the \ac{M5T}-based data rate prediction during a drive test. The dashed red lines mark performed data transmissions.}{fig:prediction_time}{-0.5cm}{0cm}{1}
This relationship is also illustrated in \Fig{\ref{fig:prediction_time}}, which shows an example trace of the temporal behavior of the \ac{M5T}-based data rate prediction during a drive test. The dependency of the data rate to the payload size can be observed by the step-wise linear component of the curve that is caused by the payload increase due to the addition of further sensor data packets with respect to the increased time.
\\
%
%
In comparison to the single-metric transmission schemes and with regard to the complex interdependencies of context parameters and data rate, it can be concluded that the data rate prediction provides a much better metric for channel quality assessment than the measured value of a single indicator. Machine learning enables the \emph{implicit consideration of hidden effects} such as \ac{TCP} slow start, channel coherence time and interdependencies between the downlink indicators into the transmission process itself.
%
%

\section{Conclusion and Future Work}

%
%
%
%
%

%
%


%
%
In this paper, we presented the machine learning-enabled transmission schemes \ac{ML-CAT} and \ac{ML-pCAT} for client-side context-aware transmission of vehicular sensor data.
%
%
Machine learning-based data rate prediction is used as a meaningful metric for scheduling the transmission time of delay-tolerant sensor data transmissions. By implicit consideration of \emph{hidden effects} such as interdependency of payload size and channel coherence time, the resulting data rate can highly be improved while simultaneously reducing the required uplink transmission power of the embedded device. The latter is a crucial factor for data sensing by energy-constraint vehicles (e.g., \acp{UAV}) and embedded systems.
The trade-off between achieved benefits and introduced delay due to local packet buffering can be controlled by different parameters and variants of the proposed probabilistic transmission scheme. 
%
%
All measurement tools and raw results of the experiments are provided in an Open Access manner in order to achieve a high level of transparency and reproducibility.
%
%
In future work, we will investigate the cross-layer interdependencies of the proposed transmission scheme. Furthermore, we aim to improve the data rate prediction by integrating knowledge about the active cell users obtained from control channel analysis. Moreover, the proposed approach will be brought together with methods for coordinated pre-aggregation of the data in vehicular crowds.
Additionally, the mobility prediction algorithms will be evaluated in their potentials for predictive steering of \ac{mmWave} pencil beams in a vehicular context.

\renewcommand{\baselinestretch}{1}
\vspace{-0.3cm}
\section*{Acknowledgment}

\footnotesize
Part of the work on this paper has been supported by Deutsche Forschungsgemeinschaft (DFG) within the Collaborative Research Center SFB 876 ``Providing Information by Resource-Constrained Analysis'', projects A1, A4, and B4.

\vspace{-0.3cm}

\bibliographystyle{IEEEtran}
\bibliography{Bibliography}

\begin{IEEEbiography}[{\includegraphics[width=1in,height=1.25in,clip,keepaspectratio]{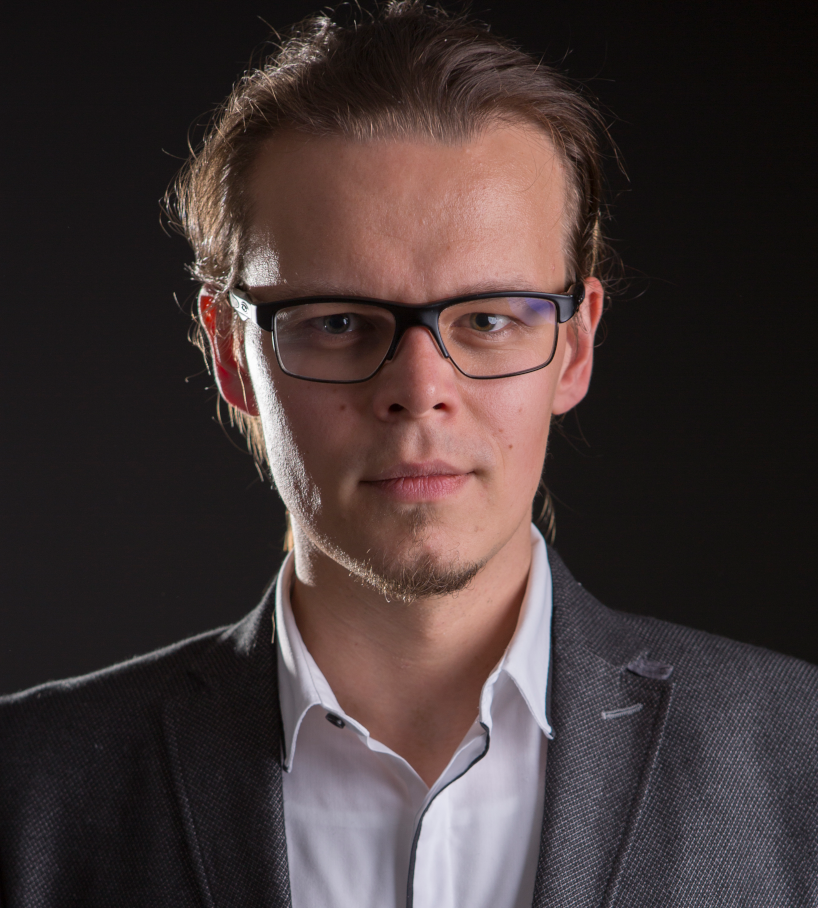}}]
	{Benjamin Sliwa}
	(S'16) received the M.Sc. degree from TU Dortmund University, Dortmund, Germany, in 2016. He is currently a Research Assistant with the Communication Networks Institute, Faculty of Electrical Engineering and Information Technology, TU Dortmund University. He is working on the Project "Analysis and Communication for Dynamic Traffic Prognosis" of the Collaborative Research Center SFB 876. His research interests include predictive and context-aware optimizations for decision processes in vehicular communication systems. 
	
	Benjamin Sliwa has been recognized with a Best Student Paper Award at IEEE VTC Spring 2018 and the 2018 IEEE Transportation Electronics Student Fellowship "For Outstanding Student Research Contributions to Machine Learning in Vehicular Communications and Intelligent Transportation Systems".
\end{IEEEbiography}

\begin{IEEEbiography}
[{\includegraphics[width=1in,height=1.25in,clip,keepaspectratio]{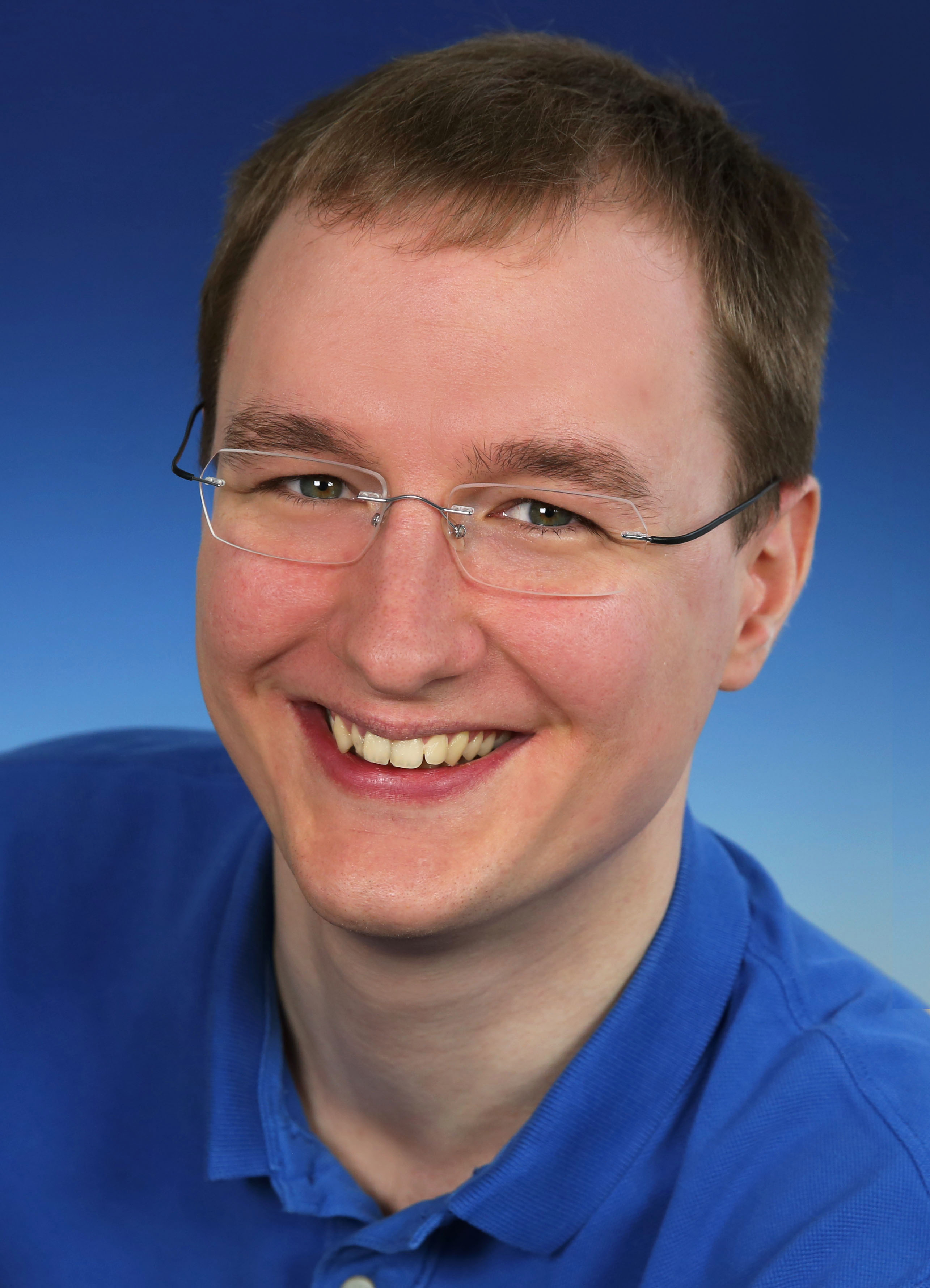}}]
{Robert Falkenberg}
(S'15) received his M.Sc. degree from TU Dortmund University, Dortmund, Germany in 2014.
He joined the Communication Network Institute, Faculty of Electrical Engineering and Information Technology, TU Dortmund University, as a Research Assistant in December 2014.
Prior to that he studied applied computer science at TU Dortmund University.
His research interests cover resource-efficient wireless communications in cellular 4G/5G and IoT networks for deeply embedded an industrial systems.
He contributes to the project ``Resource efficient and distributed platforms for integrative data analysis'' A4 of the Collaborative Research Center SFB~876.
\end{IEEEbiography}
\begin{IEEEbiography}[{\includegraphics[width=1in,height=1.25in,clip,keepaspectratio]
{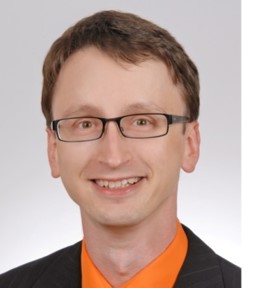}}]
{Thomas Liebig}
 is a post-doctoral researcher at the artificial
intelligence group at TU Dortmund. He works in the European H2020
project VaVeL (Variety, Veracity, VaLue: Handling the Multiplicity of
Urban Sensors http://www.vavel-project.eu) on probabilistic modeling of
spatio-temporal data, on multi-modal trip calculations and on privacy
for distributed location aware applications. He obtained his PhD from
the university of Bonn in 2013, where he studied probabilistic
pedestrian modeling from incomplete data. He has published papers in
International Conferences and Scientific Journals related to his area of
expertise (ECML/PKDD, IEEE Big Data, IEEE IE, AGILE, JAOR, JUT). He
evaluated articles for numerous journals and conferences (ICLR, ICML, NIPS,
VAST, IEEE ICDE, NEUCOM, KDD, IV, KAIS, NCAA) and serves in the program
committee of international workshops on computational transportation
sciences.
\end{IEEEbiography}

\begin{IEEEbiography}
[{\includegraphics[width=1in,height=1.25in,clip,keepaspectratio]{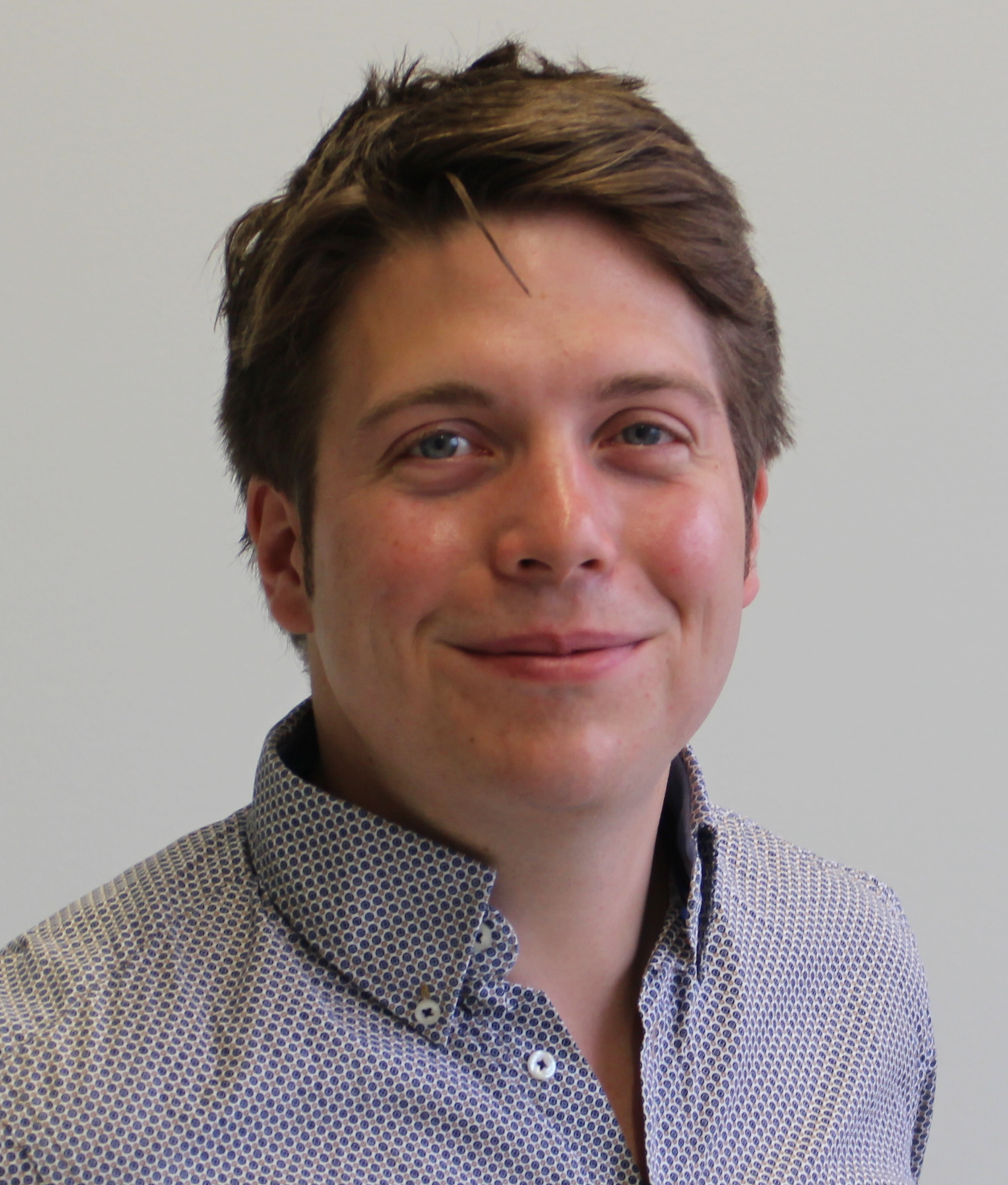}}]
{Nico Piatkowski}
studied computer science and economics at TU Dortmund in Germany where he received his PhD (Dr. rer. nat.) with summa cum laude for his work on "Exponential Families on Resource-Constrained Systems". Since 2011, Nico is a research associate at the Artificial Intelligence Unit at TU Dortmund. Nico's research is focused on machine learning, especially undirected graphical models, for resource-constrained systems. His results cover theoretical error bounds and performance guarantees for regularization, reparametrization, and numerical approximation in the context of probabilistic inference. 
\end{IEEEbiography}

\begin{IEEEbiography}
[{\includegraphics[width=1in,height=1.25in,clip,keepaspectratio]{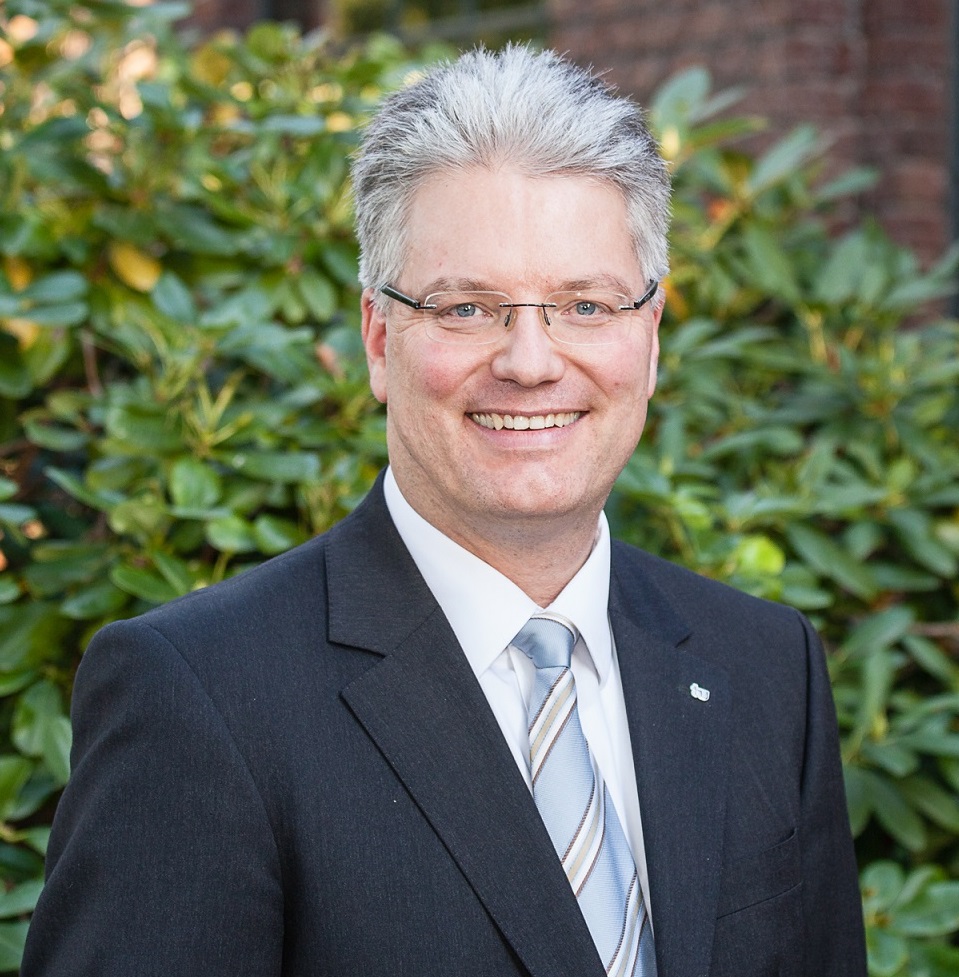}}]
{Christian Wietfeld}
(M’05–SM’12) received the Dipl.-Ing. and Dr.-Ing. degrees from RWTH Aachen University, Aachen, Germany.  He is currently a Full Professor of communication networks and the Head of the Communication Networks Institute, TU Dortmund University, Dortmund, Germany. For more than 20 years, he has been a coordinator of and a contributor to large-scale research projects on Internet-based mobile communication systems in academia (RWTH Aachen ‘92-’97, TU Dortmund since ‘05) and industry (Siemens AG ’97-’05). His current research interests include the design and performance evaluation of communication networks for cyber–physical systems in energy, transport, robotics, and emergency response.  He is the author of over 200 peer-reviewed papers and holds several patents. Dr. Wietfeld is a Co-Founder of the IEEE Global Communications Conference Workshop on Wireless Networking for Unmanned Autonomous Vehicles and member of the Technical Editor Board of the IEEE Wireless Communication Magazine. In addition to several best paper awards, he received an Outstanding Contribution award of ITU-T for his work on the standardization of next-generation mobile network architectures.

\end{IEEEbiography}

\end{document}